\documentclass[iop,apj,twocolappendix]{emulateapj}
\usepackage{apjfonts}
\usepackage{multirow}
\usepackage{textcomp}
\usepackage{amsmath}
\usepackage{acronym}
\usepackage[colorlinks,citecolor=blue,linkcolor=blue,urlcolor=blue]{hyperref}
\newacro{SNR}[S/N]{signal-to-noise ratio}
\newacro{RMS}[RMS]{root mean square}

\begin{document}

\title{The First Two Years of Electromagnetic Follow\nobreakdashes-Up with Advanced LIGO and Virgo}

\slugcomment{The Astrophysical Journal, 795:105}
\received{2014 April 23}
\accepted{2014 September 4}
\published{2014 October 17}

\author{Leo P. Singer\altaffilmark{1}}
\email{lsinger@caltech.edu}
\author{Larry R. Price\altaffilmark{1}}
\author{Ben Farr\altaffilmark{2,3}}
\author{Alex L. Urban\altaffilmark{4}}
\author{Chris Pankow\altaffilmark{4}}
\author{Salvatore Vitale\altaffilmark{5}}
\author{John Veitch\altaffilmark{6,3}}
\author{Will M. Farr\altaffilmark{3}}
\author{Chad Hanna\altaffilmark{7,8}}
\author{Kipp Cannon\altaffilmark{9}}
\author{Tom Downes\altaffilmark{4}}
\author{Philip Graff\altaffilmark{10}}
\author{Carl-Johan Haster\altaffilmark{3}}
\author{Ilya Mandel\altaffilmark{3}}
\author{Trevor Sidery\altaffilmark{3}}
\author{Alberto Vecchio\altaffilmark{3}}

\altaffiltext{1}{LIGO Laboratory, California Institute of Technology, Pasadena, CA 91125, USA}
\altaffiltext{2}{Department of Physics and Astronomy \& Center for Interdisciplinary Exploration and Research in Astrophysics (CIERA), Northwestern University, Evanston, IL 60208, USA}
\altaffiltext{3}{School of Physics and Astronomy, University of Birmingham, Birmingham, B15 2TT, UK}
\altaffiltext{4}{Leonard E. Parker Center for Gravitation, Cosmology, and Astrophysics, University of Wisconsin\nobreakdashes--Milwaukee, Milwaukee, WI 53201, USA}
\altaffiltext{5}{Massachusetts Institute of Technology, 185 Albany Street, Cambridge, MA 02139, USA}
\altaffiltext{6}{Nikhef, Science Park 105, Amsterdam 1098XG, The Netherlands}
\altaffiltext{7}{Perimeter Institute for Theoretical Physics, Ontario N2L 2Y5, Canada}
\altaffiltext{8}{The Pennsylvania State University, University Park, PA 16802, USA}
\altaffiltext{9}{Canadian Institute for Theoretical Astrophysics, University of Toronto, Toronto, Ontario, M5S 3H8, Canada}
\altaffiltext{10}{NASA Goddard Space Flight Center, Greenbelt, MD, USA}

\shorttitle{Advanced LIGO and Virgo: The first two years}
\shortauthors{Singer et al.}


\begin{abstract}
We anticipate the first direct detections of gravitational waves (GWs) with Advanced LIGO and Virgo later this decade. Though this groundbreaking technical achievement will be its own reward, a still greater prize could be observations of compact binary mergers in both gravitational and electromagnetic channels simultaneously. During Advanced LIGO and Virgo's first two years of operation, 2015 through 2016, we expect the global GW detector array to improve in sensitivity and livetime and expand from two to three detectors. We model the detection rate and the sky localization accuracy for binary neutron star (BNS) mergers across this transition. We have analyzed a large, astrophysically motivated source population using real\nobreakdashes-time detection and sky localization codes and higher\nobreakdashes-latency parameter estimation codes that have been expressly built for operation in the Advanced LIGO/Virgo era. We show that for most BNS events the rapid sky localization, available about a minute after a detection, is as accurate as the full parameter estimation. We demonstrate that Advanced Virgo will play an important role in sky localization, even though it is anticipated to come online with only one\nobreakdashes-third as much sensitivity as the Advanced LIGO detectors. We find that the median 90\% confidence region shrinks from $\sim$500\,deg$^2$ in 2015 to $\sim$200\,deg$^2$ in 2016. A few distinct scenarios for the first LIGO/Virgo detections emerge from our simulations.
\end{abstract}

\section{Introduction}

We expect this decade to bring the first direct detection of gravitational waves (GWs) from compact objects. The LIGO and Virgo detectors are being rebuilt with redesigned mirror suspensions, bigger optics, novel optical coatings, and higher laser power~\citep{aLIGO, aVirgo}. In their final configuration, Advanced LIGO and Virgo are expected to reach $\sim10$ times further into the local universe than their initial configurations did. The best\nobreakdashes-understood sources for LIGO and Virgo are binary neutron star (BNS) mergers. They also offer a multitude of plausible electromagnetic (EM) counterparts~\citep{MostPromisingEMCounterpart} including collimated short\nobreakdashes-hard gamma\nobreakdashes-ray bursts (short GRBs; see for example \citealt{1986ApJ...308L..43P,1989Natur.340..126E,1992ApJ...395L..83N,2011ApJ...732L...6R}) and X\nobreakdashes-ray/optical afterglows, near\nobreakdashes-infrared kilonovae \citep[viewable from all angles;][etc.]{kilonova, BarnesKasenKilonovaOpacities}, and late\nobreakdashes-time radio emission~\citep{NakarPiranRadioFlares,PiranNakarRosswogEMSignals}. Yet, typically poor GW localizations of $\gtrsim 100\text{\,deg}^2$ will present formidable challenges to observers hunting for their EM counterparts.

Several planned optical astronomy projects with a range of fields of view and apertures are preparing to pursue these elusive events. These include the Zwicky Transient Facility~\citep{ZTF}, PanSTARRS\footnote{\url{http://pan-starrs.ifa.hawaii.edu/public/}}, BlackGEM\footnote{\url{https://www.astro.ru.nl/wiki/research/blackgemarray}}, and LSST~\citep{LSST}, to name a few. Advanced LIGO is scheduled to start taking data in 2015~\citep{LIGOObservingScenarios}. Preparations for joint EM and GW observations require a complete understanding of when and how well localized the first GW detections will be. Plausible scenarios for the evolution of the configuration and sensitivity of the worldwide GW detector network as it evolves from 2015 through 2022, as well as rough estimates of sky localization area, are outlined in \citet{LIGOObservingScenarios}.

To provide a more realistic and complete picture, we have conducted Monte Carlo simulations of the 2015 and 2016 detector network configurations, probing the transition from two to three detectors as Advanced Virgo is scheduled to begin science operation. Prior work has focused on various aspects of position reconstruction with advanced GW detectors \citep{FairhurstTriangulation,WenLocalizationAdvancedLIGO,FairhurstLocalizationAdvancedLIGO,2011PhRvD..84j4020V,RodriguezBasicParameterEstimation,NissankeLocalization,NissankeKasliwalEMCounterparts,KasliwalTwoDetectors,Grover:2013,SiderySkyLocalizationComparison}, but ours is the first to bring together a large astrophysically motivated population, an educated guess about the detector commissioning timetable, a realistic \ac{SNR} distribution, and the Advanced LIGO/Virgo data analysis pipeline itself.

We have simulated hundreds of GW events, recovered them with a real\nobreakdashes-time detection pipeline, and generated sky maps using both real\nobreakdashes-time and thorough off\nobreakdashes-line parameter estimation codes that will be operating in 2015 and beyond. This study contains some of the first results with \textsc{bayestar}, a rapid Bayesian position reconstruction code that will produce accurate sky maps less than a minute after any BNS merger detection. The \textsc{lalinference\_mcmc} \citep{2008ApJ...688L..61V,Raymond:2009}, \textsc{lalinference\_nest} \citep{LALINFERENCE_NEST}, and \textsc{lalinference\_bambi} \citep{BAMBI,SKYNET} stochastic samplers were also used to follow up a subset of detected GW events. Though these analyses are significantly more computationally costly than \textsc{bayestar}, taking hours to days, they can provide improved sky location estimates when the GW signal is very weak in one detector, and also yield not just sky localization but the full multidimensional probability distribution describing the parameters of a circularized compact binary merger. All four algorithms are part of the \textsc{lalinference} library \citep{S6PE}, developed specifically for estimating the parameters of GW sources from ground-based detectors. Together, these analyses will be able to provide sky localizations on time scales that enable searching for all expected electromagnetic counterparts of compact binary mergers (except the GRB itself).

With the benefit of a much larger sample size, important features of the 2015 and 2016 configurations come into focus. First, we find that even in 2015 when only the two LIGO detectors are operating (or in 2016 during periods when the Virgo detector is not in science mode), there is at least a 60\% chance of encountering the source upon searching an area of about 200\,deg$^2$. Second, many of these two\nobreakdashes-detector events will not be localized to a single simply connected region in the sky. We elucidate two nearly degenerate sky locations, separated by 180$^\circ$, that arise when only the two LIGO detectors are operating. When a GW source falls within this degeneracy, its sky map will consist of two diametrically opposed islands of probability. Third, in our simulations, we add a third detector, Advanced Virgo, in 2016. Even though at that time Virgo is anticipated to be only one\nobreakdashes-third as sensitive as the other two detectors due to differing LIGO and Virgo commissioning timetables, we find that coherence with the signal in Virgo generally breaks the previously mentioned degeneracy and shrinks areas to a third of what they were with two detectors. Fourth and most importantly, a picture of a typical early Advanced LIGO event emerges, with most occurring in a limited range of Earth\nobreakdashes-fixed locations, and most sky maps broadly fitting a small number of specific morphologies. 

\section{Sources and Sensitivity}

BNS systems are the most promising and best understood targets for joint GW and EM detection. Though rate estimates remain uncertain, ranging from 0.01 to 10\,Mpc$^{-3}$\,Myr$^{-1}$, we choose to work with the ``realistic'' rate obtained from \citet{LIGORates} of 1\,Mpc$^{-3}$\,Myr$^{-1}$. This rate leads to a GW detection rate of 40\,yr$^{-1}$ at final Advanced LIGO design sensitivity. Some mergers of neutron star--black hole binaries (NSBHs) are also promising sources of GW and EM emission. Two Galactic high\nobreakdashes-mass X\nobreakdashes-ray binaries (HMXBs) have been identified as possible NSBH progenitors \citep{CygX1NSBHBBH,CygX3NSBHBBH}. From these can be extrapolated a lower bound on the GW detection rate of at least 0.1\,yr$^{-1}$ at Advanced LIGO's final design sensitivity, although rates comparable to BNS detections are empirically plausible. Black holes in binaries may possess large spins, causing precession during the inspiral. Precession\nobreakdashes-altered phase evolution can aid in parameter estimation~\citep{2008CQGra..25r4011V,2008ApJ...688L..61V,Harry:2013tca,Nitz:2013mxa,Raymond:2009}, but models of waveforms suitable for rapid detection and parameter estimation are still under active development \citep{Blackman:2014maa,Hannam:2013waveform,Taracchini:2013}. As for the binary black hole mergers detectable by Advanced LIGO and Virgo, there are currently no compelling mechanisms for electromagnetic counterparts associated with them. We therefore restrict our attention to BNS mergers, because they have the best understood rates, GW signal models, and data analysis methods.

\subsection{Measures of detector sensitivity}

The sensitivity of a single GW detector is customarily described by the horizon distance, or the maximum distance at which a particular source would create a signal with a maximum fiducial single\nobreakdashes-detector \ac{SNR}, $\rho$.\footnote{Even at its final design sensitivity, Advanced LIGO's range for BNS mergers is only 200\,Mpc or $z = 0.045$ \citep[assuming the \emph{WMAP} nine\nobreakdashes-year $\Lambda$CDM cosmology;][]{WMAP9}. Because of the small distances considered in this study, we do not distinguish between different distance measures, nor do our gravitational waveforms contain any factors of $(1+z).$} It is given by
\begin{equation}
    \label{eq:horizon-distance}
    d_\mathrm{H} \approx \frac{G^{5/6}M^{1/3}\mu^{1/2}}{c^{3/2}\pi^{2/3}\rho}\sqrt{\frac{5}{6} \int_{f_1}^{f_2} \frac{f^{-7/3}}{S(f)}\,\mathrm{d}f}
\end{equation}
where $G$ is Newton's gravitational constant, $c$ is the speed of light, $M$ the sum of the component masses, $\mu$ the reduced mass, $f^{-7/3}$ the approximate power spectral density (PSD) of the inspiral signal, and $S(f)$ the PSD of the detector's noise. The lower integration limit $f_1$ is the low\nobreakdashes-frequency extent of the detector's sensitive band. For the Advanced LIGO and Virgo detectors, ultimately limited at low frequency by ground motion \citep{GWDetectionLaserInterferometry}, we take $f_1 = 10$\,Hz. Using a typical value of the detector sensitivity $S(100\,\text{Hz}) = 10^{-46}\,\text{Hz}^{-1}$, we can write Equation~(\ref{eq:horizon-distance}) as a scaling law,
\begin{multline}
    \label{eq:horizon-distance-scaling}
    d_\mathrm{H} \approx
        72.5\,\text{Mpc}
        \left(\frac{M}{M_\odot}\right)^{1/3}
        \left(\frac{\mu}{M_\odot}\right)^{1/2}
        \left(\frac{1}{\rho}\right) \\
    \cdot
        \left[
            \int_{\frac{f_1}{\text{Hz}}}^{\frac{f_2}{\text{Hz}}}
            \left(\frac{f}{100\,\text{Hz}}\right)^{-7/3} 
            \left(\frac{10^{-46}\,\text{Hz}^{-1}}{S(f)}\right)
            d\left(\frac{f}{\text{Hz}}\right) \right]^{1/2}.
\end{multline}
For BNS masses, the inspiral ends with a merger and black hole ring down well outside LIGO's most sensitive band. A reasonable approximation is to simply truncate the \ac{SNR} integration at the last stable orbit of a Schwarzschild black hole with the same total mass \citep{maggiore2008gravitational},
\begin{equation}
    \label{eq:f-lso}
    f_2 \approx (4400\text{\,Hz}) \frac{M_\odot}{M}.
\end{equation}
Usually, $\rho=8$ is assumed because $\rho=8$ signals in two detectors (for a root\nobreakdashes-sum\nobreakdashes-squared network \ac{SNR} of $\rho_\mathrm{net} = 8\sqrt{2} = 11.3$) is nearly adequate for a confident detection (see discussion of detection thresholds in Section~\ref{sec:detection-and-position-reconstruction}). Another measure of sensitivity is the BNS range $d_\mathrm{R}$, the volume-, direction-, and orientation\nobreakdashes-averaged distance of a source with $\rho \geq 8$, drawn from a homogeneous population. Due to the directional sensitivity or antenna pattern of interferometric detectors, the range is a factor of 2.26 smaller than the horizon distance for the same \ac{SNR} threshold. See also \citet{FINDCHIRP,S6Sensitivity}.

\subsection{Observing Scenarios}

\begin{figure*}
    \begin{minipage}[b]{3.5in}
        \begin{center}
            \includegraphics{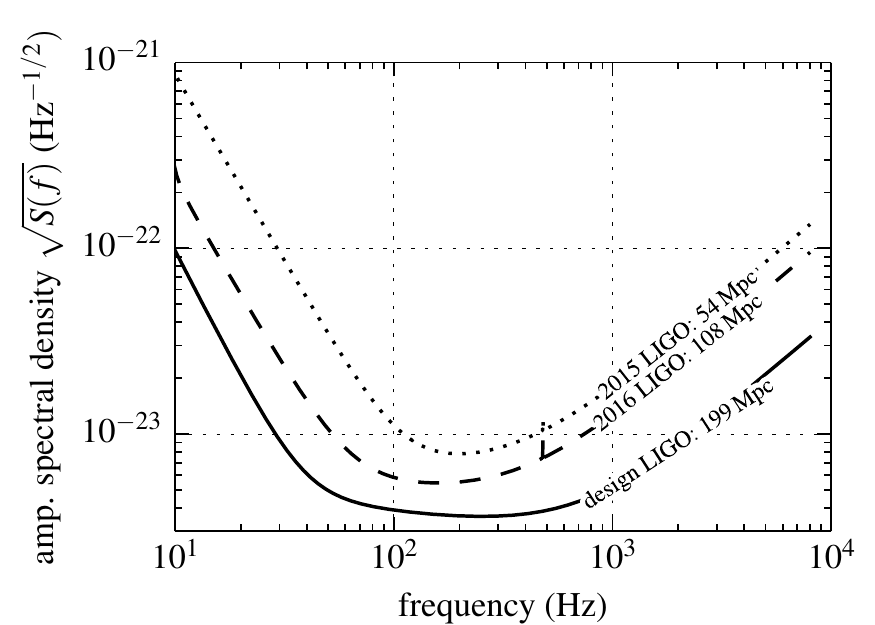}

            (a) LIGO
        \end{center}
    \end{minipage}
    \begin{minipage}[b]{3.5in}
        \begin{center}
            \includegraphics{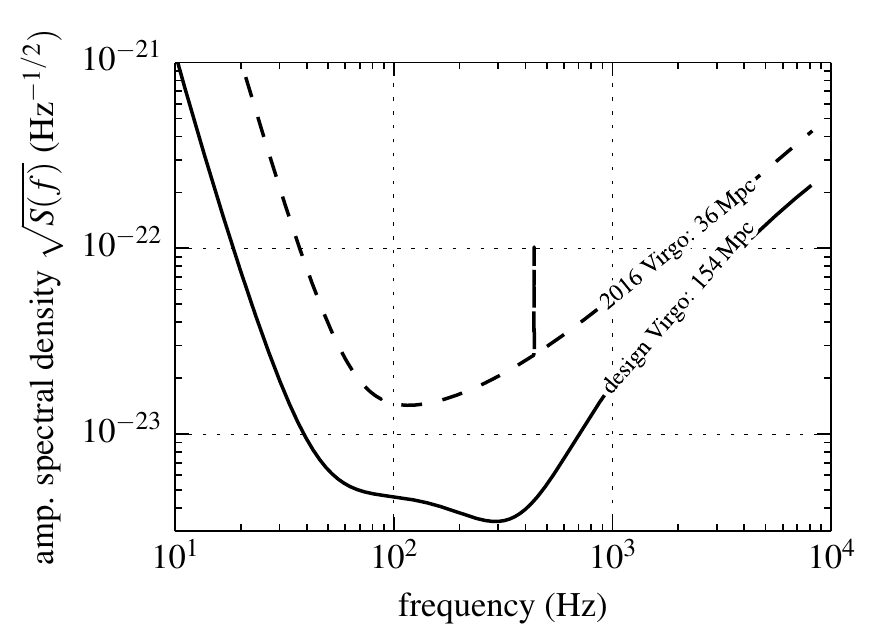}

            (b) Virgo
        \end{center}
    \end{minipage}
    \caption{\label{fig:psds}Model detector noise amplitude spectral density curves. The LIGO 2015, 2016, and final design noise curves are shown in the left panel and the Virgo 2016 and final design noise curves in the right panel. The averaged $\rho=8$ range $d_\mathrm{R}$ for $(1.4, 1.4)$\,$M_\odot$ BNS mergers is given for each detector.}
\end{figure*}

\citet{LIGOObservingScenarios} outline five observing scenarios representing the evolving configuration and capability of the Advanced GW detector array, from the first observing run in 2015, to achieving final design sensitivity in 2019, to adding a fourth detector at design sensitivity by 2022. In this study, we focus on the first two epochs. The first, in 2015, is envisioned as a three\nobreakdashes-month science run. LIGO Hanford (H) and LIGO Livingston (L) Observatories are operating with an averaged $(1.4, 1.4)\,M_\odot$ BNS range between 40~and~80\,Mpc. The second, in 2016\nobreakdashes--2017, is a six\nobreakdashes-month run with H and L operating between 80~and~120\,Mpc and the addition of Advanced Virgo (V) with a range between 20~and~60\,Mpc. For each configuration, we used model noise PSD curves in the middle of the ranges in \citet{LIGOObservingScenarios}, plotted in Figure~\ref{fig:psds}. For H and L, we used the ``early'' and ``mid'' noise curves from \citet{EarlyAdvancedLIGONoiseCurves} for the 2015 and 2016 scenarios respectively. For V in 2016, we used the geometric mean of the high and low curves of \citet{LIGOObservingScenarios}. Final LIGO and Virgo design sensitivity is several steps further in the commissioning schedule than we consider in this paper.

\subsection{Simulated Waveforms}

For each of the two scenarios we made synthetic detector streams by placing post\nobreakdashes-Newtonian inspiral signals into two months of colored Gaussian noise. We used ``SpinTaylorT4'' waveforms, employing the TaylorT4 approximant and accurate to 3.5PN order in phase and 1.5PN order in amplitude \citep{SpinTaylorT4, SpinTaylorT4Erratum, taylorf2}.\footnote{There is a C language implementation as the function \\ \mbox{\texttt{XLALSimInspiralSpinTaylorT4}} in \textsc{lalsimulation}. See \\ Acknowledgments and Appendix.} There was an average waiting time of $\approx$100\,s between coalescences. At any given time, one BNS inspiral signal was entering LIGO's sensitive band while another binary was merging, but both signals were cleanly separated due to their extreme narrowness in time\nobreakdashes-frequency space. The PSD estimation used enough averaging that it was unaffected by the overlapping signals. Component masses were distributed uniformly between 1.2~and~1.6\,$M_\odot$, bracketing measured masses for components of known BNS systems as well as the 1\nobreakdashes-$\sigma$ intervals of the intrinsic mass distributions inferred for a variety of NS formation channels~\citep{NeutronStarMass1, NeutronStarMass2}.

We gave each NS a randomly oriented spin with a maximum magnitude of $\chi = c\left|\mathbf{S}\right|/Gm^2 \leq 0.05$, where $\mathbf{S}$ is the star's spin angular momentum and $m$ is its mass. This range includes the most rapidly rotating pulsar that has been found in a binary, PSR\,J0737\nobreakdashes-3039A \citep{2003Natur.426..531B,DetectingBNSSystemsWithSpin}. However, the fastest\nobreakdashes-spinning millisecond pulsar, PSR\,J1748\nobreakdashes-2446ad \citep{FastestSpinningMillisecondPulsar}, has a dimensionless spin parameter of $\sim$0.4, and the theoretical evolutionary limits on NS spin\nobreakdashes-up in BNS systems are uncertain.

\subsection{Sensitivity to Assumptions}

The total detection rate depends on some of these assumptions, and in particular is sensitive to the assumed NS mass distribution. As can be seen from Equation~(\ref{eq:horizon-distance}), binaries with the greatest and most symmetric component masses can be detected to the farthest distance. According to Equation~(\ref{eq:f-lso}), for BNS systems the merger always occurs at kHz frequencies, on the upward $S(f) \propto f^2$ slope of the noise noise curves in Figure~\ref{fig:psds} in the regime dominated by photon shot noise~\citep{QuantumNoiseSecondGeneration,GWDetectionLaserInterferometry}. As a result, the integral in Equation~(\ref{eq:horizon-distance}) depends only weakly on masses. For equal component masses, the horizon distance scales as $d_\mathrm{H} \propto {m_\mathrm{NS}}^{5/6}$, so the detection rate scales rapidly with mass as $\dot{N} \propto {d_\mathrm{H}}^3 \propto {m_\mathrm{NS}}^{2.5}$.

The normalized distribution of sky localization areas depends only weakly on the distribution of NS masses. \citet{FairhurstTriangulation} computes the approximate scaling of sky localization area by considering the Fisher information associated with time of arrival measurement. Valid for moderately high \ac{SNR}, the \ac{RMS} uncertainty in the time of arrival in a given detector is
\begin{equation}
    \label{eq:sigmat}
    \sigma_t = \frac{1}{2 \pi \rho \sqrt{\overline{f^2} - \overline{f}^2}}
\end{equation}
where $\overline{f} = \overline{f^1}$, and $\overline{f^k}$ is the $k$th moment of frequency, weighted by the signal to noise per unit frequency,
\begin{equation}
    \overline{f^k} \approx \left[ \int_{f_1}^{f_2} \frac{|h(f)|^2 f^k}{S(f)}\,\mathrm{d}f \right] \left[ \int_{f_1}^{f_2} \frac{|h(f)|^2}{S(f)}\,\mathrm{d}f \right]^{-1}.
\end{equation}
As in Equation~(\ref{eq:horizon-distance}), we can substitute the approximate inspiral signal spectrum $|h(f)|^2 \propto f^{-7/3}$. The areas then scale as the product of the timing uncertainty in individual detectors, or as simply the square of Equation~(\ref{eq:sigmat}) for a network of detectors with similar (up to proportionality) noise PSDs. As $m_\mathrm{NS}$ varies from 1 to 2\,$M_\odot$, the upper limit of integration $f_2$ given by Equation~(\ref{eq:f-lso}) changes somewhat, but areas change by a factor of less than 1.5. (See also \citealt{Grover:2013} for scaling of sky localization area with mass).

Introducing faster NS spins would result in smaller sky localization areas, since orbital precession can aid in breaking GW parameter estimation degeneracies \citep{Raymond:2009}. However, rapid spins could require more exotic BNS formation channels, and certainly would require using more sophisticated and more computationally expensive GW waveforms for parameter estimation.

\subsection{Source Locations}

Source locations were random and isotropic, and uniform in distance cubed. The source distribution was cut off at the $\rho=5$, $(1.6, 1.6)\,M_\odot$ horizon distance, far enough away that the selection of detected binaries was determined primarily by the sensitivity of the instruments and the detection pipeline, not by the artificial distance boundary.

\subsection{Duty Cycle}

Following \citet{LIGOObservingScenarios}, we assumed that each detector had an independent and random $d=80\%$ duty cycle. In the 2015, HL configuration, this implies that both detectors are in operation $d^2=64\%$ of the time. In 2016, there are three detectors operating $d^3 = 51.2\%$ of the time and each of three pairs operating $d^2(1-d)=12.8\%$ of the time. We do not simulate any of the time when there are one or zero detectors operating, but instead fold this into conversion factors from our Monte Carlo counts to detection rates.

\section{Detection and Position Reconstruction}
\label{sec:detection-and-position-reconstruction}

\begin{figure*}
    \includegraphics[width=1.1\textwidth]{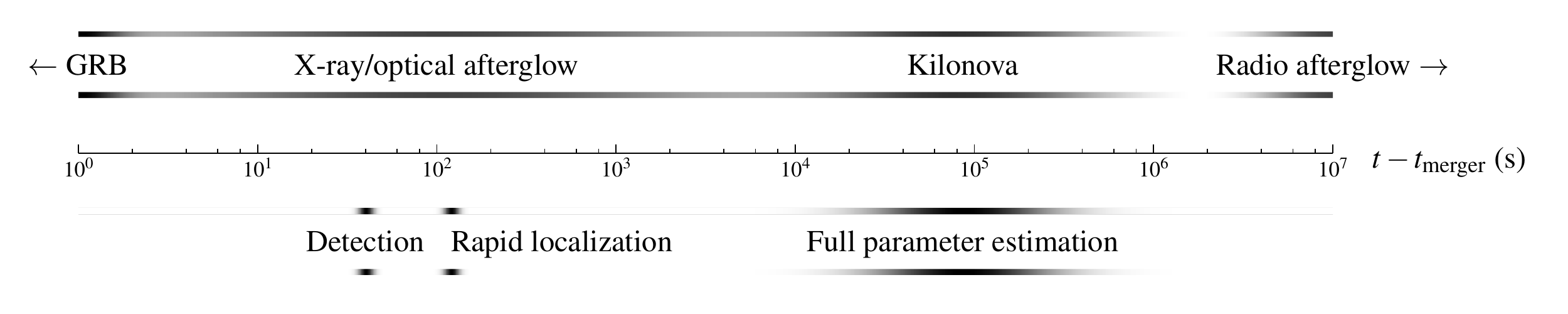}
    \caption{\label{fig:timeline}Rough timeline of compact binary merger electromagnetic emissions in relation to the timescale of the Advanced LIGO/Virgo analysis described in this paper. The time axis measures seconds after the merger.}
\end{figure*}

Searches for GWs from compact binaries \citep{FINDCHIRP,ihope} employ banks of matched filters, in the data from all of the detectors are convolved with an array of template waveforms. The output of each filter is the instantaneous \ac{SNR} with respect to that template in that detector. An excursion above a threshold \ac{SNR} in two or more detectors with exactly the same binary parameters and within approximately one light\nobreakdashes-travel time between detectors is considered a coincidence. Coincidences may be accidental, due to chance noise fluctuations or, in real GW data streams, environmental disturbances and instrument glitches. Coincidences with sufficiently high $\rho_\mathrm{net}$ (root\nobreakdashes-sum\nobreakdashes-square of the \ac{SNR} in the individual detectors) are considered detection candidates. A $\chi^2$ statistic is used to aid in separating the true, astrophysical signals from accidental coincidences or false positives~\citep{AllenChiSq, HannaThesis, Cannon:2012zt}.

Offline inspiral searches used in past LIGO/Virgo science runs will be computationally strained in Advanced LIGO/Virgo due to denser template banks and BNS signals that remain in band for up to $\sim 10^3$\,s. To address these issues and achieve latencies of $\lesssim 1$\,minutes, a rapid matched\nobreakdashes-filter detection pipeline called \textsc{gstlal\_inspiral}~\citep{Cannon:2011vi} has been developed. To mimic Advanced LIGO/Virgo observations as closely as possible, we used \textsc{gstlal\_inspiral} to extract simulated detection candidates from our two\nobreakdashes-month data streams.

\subsection{Template Waveforms}

The templates were constructed from a frequency domain, post\nobreakdashes-Newtonian model describing the inspiral of two compact objects, accurate to 3.5 post\nobreakdashes-Newtonian order in phase and Newtonian order in amplitude \citep{taylorf2}.\footnote{These are in \textsc{lalsimulation} as the function \\ \mbox{\texttt{XLALSimInspiralTaylorF2}}. See acknowledgements and Appendix.} These waveforms neglect spins entirely. This is known to have a minimal impact on detection efficiency for BNS sources with low spins \citep{DetectingBNSSystemsWithSpin}. These waveforms are adequate for recovering the weakly spinning simulated signals that we placed into the data stream.

\subsection{Detection Threshold}

In our study, we imposed a single-detector threshold \ac{SNR} of 4. A simulated signal is then considered to be detected by \textsc{gstlal\_inspiral} if it gives rise to a coincidence with sufficiently low false alarm probability as estimated from the \ac{SNR} and $\chi^2$ values. We follow the lead of \citet{LIGOObservingScenarios} in adopting a false alarm rate (FAR) threshold of $\textsc{FAR} \leq 10^{-2}$\,yr$^{-1}$. \citet{LIGOObservingScenarios} claim that in data of similar quality to previous LIGO/Virgo science runs, this FAR threshold corresponds to a network \ac{SNR} threshold of $\rho_\mathrm{net} \geq 12$. Since our data is Gaussian and perfectly free of glitches, to obtain easily reproducible results we imposed a second explicit detection cut of $\rho_\mathrm{net} \geq 12$. We find that our joint threshold on FAR and \ac{SNR} differs negligibly from a threshold on \ac{SNR} alone. Because any given simulated signal will cause multiple coincidences at slightly different masses and arrival times, for each simulated signal we keep only the matching candidate with the lowest \ac{SNR}.

\subsection{Sky Localization and Parameter Estimation}

All detection candidates are followed up with rapid sky localization by \textsc{bayestar} and a subset were followed up with full parameter estimation by the \textsc{lalinference\_mcmc}/\textsc{nest}/\textsc{bambi} stochastic samplers. The three different stochastic samplers all use the same likelihood, but serve as useful cross\nobreakdashes-verification. Both \textsc{bayestar} and the three stochastic samplers are coherent (exploiting the phase consistency across all detectors) and Bayesian (making use of both the GW observations and prior distributions over the source parameters). They differ primarily in their input data.

\textsc{bayestar}'s likelihood function depends on only the information associated with the triggers comprising a coincidence: the times, phases, and amplitudes on arrival at each of the detectors. \textsc{bayestar} exploits the leading\nobreakdashes-order independence of errors in the extrinsic and intrinsic parameters by holding the masses fixed at the values estimated by the detection pipeline. Marginalized posterior distributions for the sky positions of sources are produced by numerically integrating the posterior in each pixel of the sky map. Because \textsc{bayestar}'s analysis explores only a small sector of the full parameter space, never performs costly evaluations of the post-Newtonian GW waveforms, and uses highly tuned standard numerical quadrature techniques, it takes well under a minute (see Figure~\ref{fig:timeline}).

On the other hand, the likelihood function used for the stochastic samplers depends on the full GW data, and is the combination of independent Gaussian distributions for the residual in each frequency bin after model subtraction. Posterior distributions for the sky position are produced by sampling the full parameter space of the signal model, then marginalizing over all parameters but the sky location. This method requires the generation of a model waveform for each sample in parameter space, making it far more expensive than the \textsc{bayestar} approach, but independent of the methods and models used for detection. Most importantly, intrinsic parameters (including spins) can be estimated using these higher-latency methods. For the purposes of this study, parameter estimation used the same frequency\nobreakdashes-domain, non\nobreakdashes-spinning waveform approximant as the detection pipeline. Analyses that account for the spin of the compact objects are more costly, currently taking weeks instead of days to complete, and will be the subject of a future study.

\section{Results}

\begin{figure*}
    \begin{minipage}[b]{3.5in}
        \begin{center}
            \includegraphics{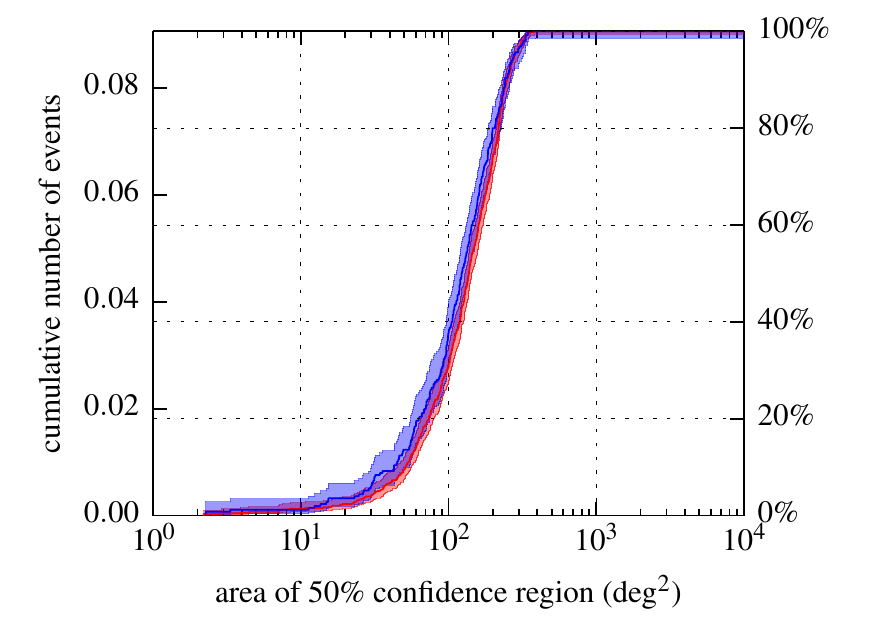}

            (a) 2015, HL
        \end{center}
    \end{minipage}
    \begin{minipage}[b]{3.5in}
        \begin{center}
            \includegraphics{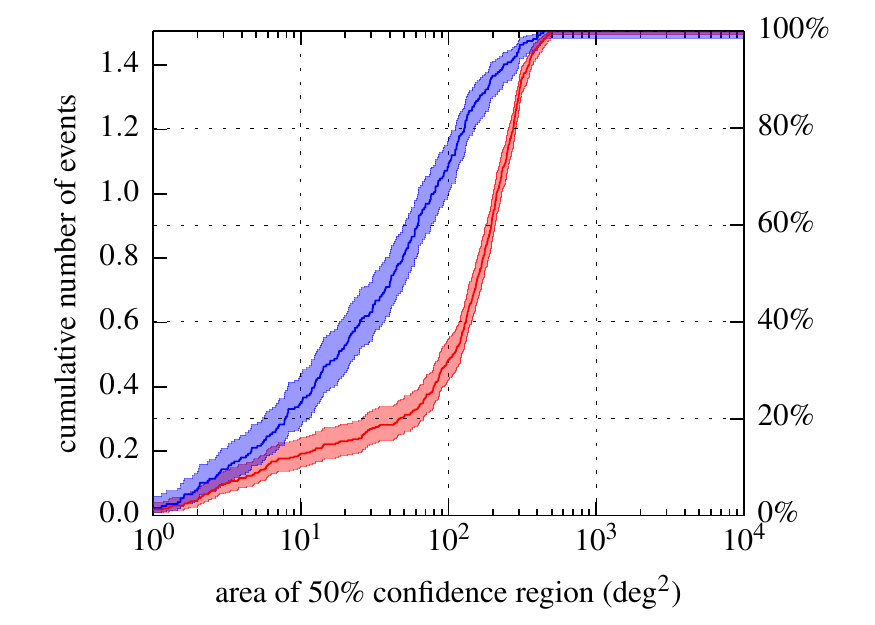}

            (b) 2016, HLV
        \end{center}
    \end{minipage}

    \begin{minipage}[b]{3.5in}
        \begin{center}
            \includegraphics{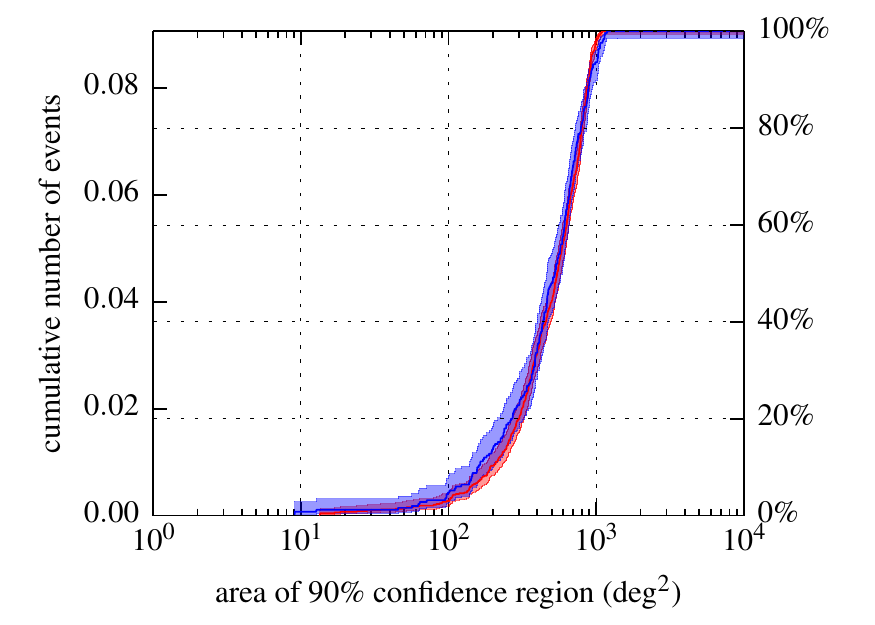}

            (c) 2015, HL
        \end{center}
    \end{minipage}
    \begin{minipage}[b]{3.5in}
        \begin{center}
            \includegraphics{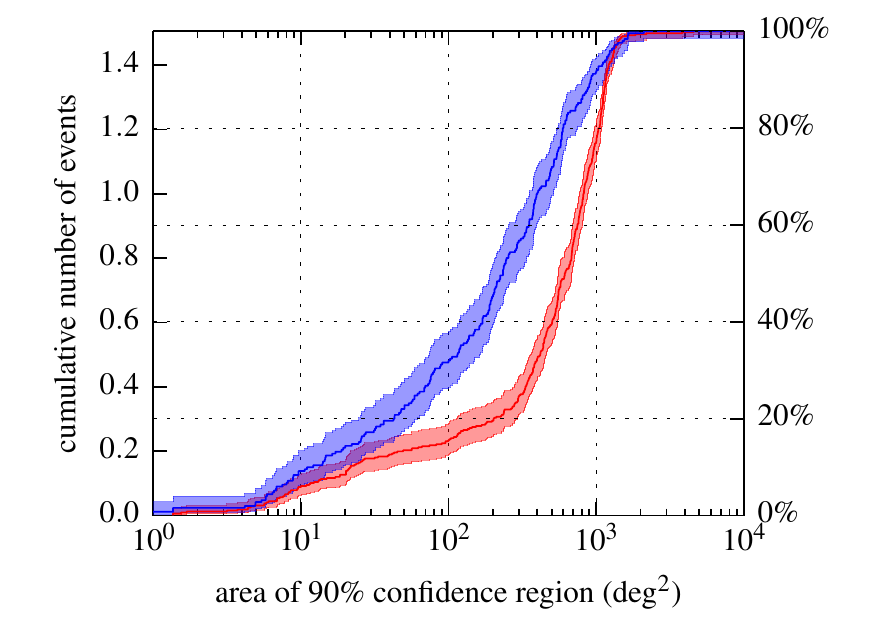}

            (d) 2016, HLV
        \end{center}
    \end{minipage}

    \begin{minipage}[b]{3.5in}
        \begin{center}
            \includegraphics{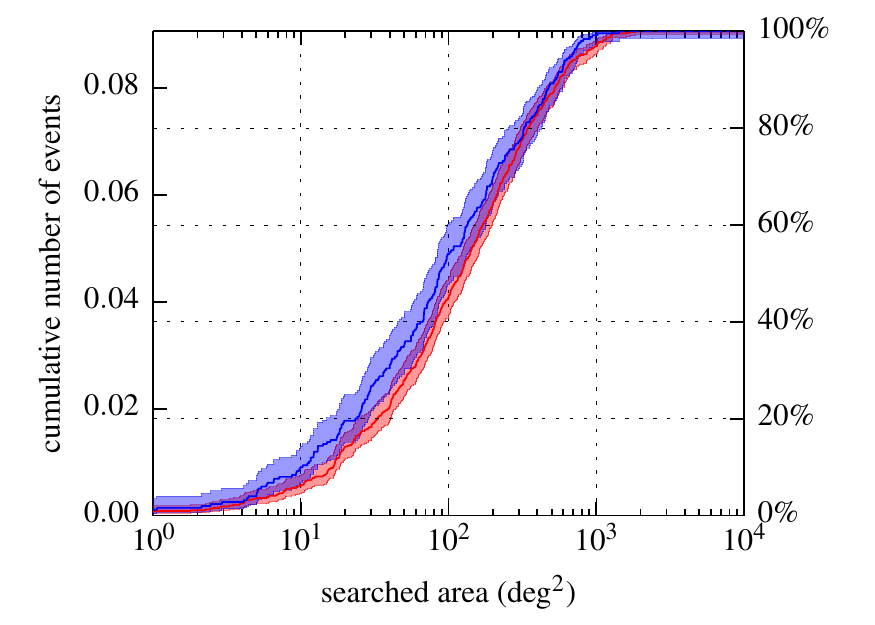}

            (e) 2015, HL
        \end{center}
    \end{minipage}
    \begin{minipage}[b]{3.5in}
        \begin{center}
            \includegraphics{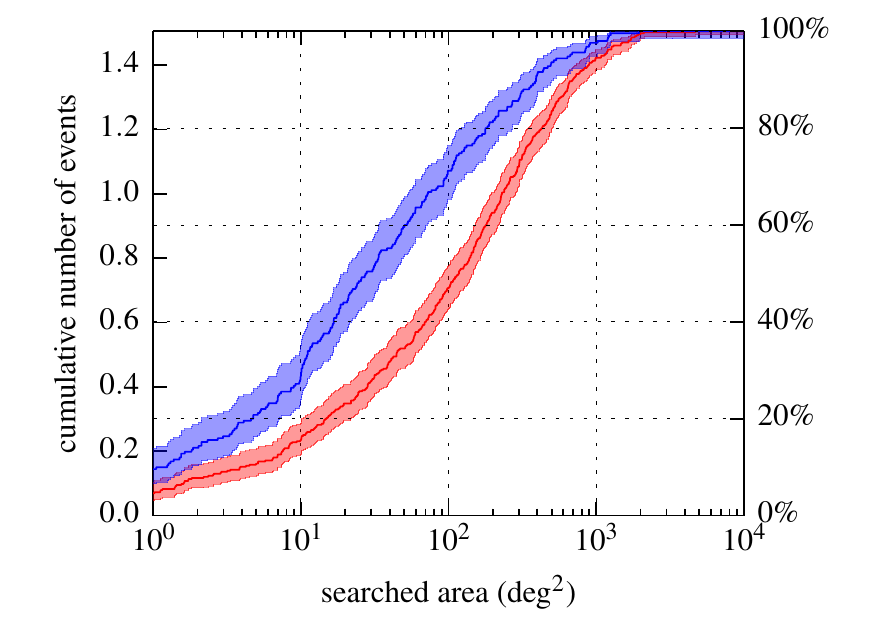}

            (f) 2016, HLV
        \end{center}
    \end{minipage}
    \caption{\label{fig:area_hist}Cumulative histogram of sky localization areas in the 2015 (HL) and 2016 (HLV) scenarios. Plots in the left column (a,~c,~e) refer to the 2015 configuration and in the right column (b,~d,~f) to the 2016 configuration. The first row (a,~b) shows the area of the 50\% confidence region, the second row (c,~d) shows the 90\% confidence region, and the third row (e,~f) shows the ``searched area,'' the area of the smallest confidence region containing the true location of the source. The red lines comprise all detections and their sky maps produced with \textsc{bayestar}, and the blue lines represent sky maps for the random subsample of 250 detections analyzed with \textsc{lalinference\_nest}/\textsc{mcmc}. The light shaded region encloses a 95\% confidence interval accounting for sampling errors \citep[computed from the quantiles of a beta distribution;][]{BinomialConfidenceIntervalsAstronomy}. The left axes show the number of detections localized within a given area assuming the ``realistic'' BNS merger rates from \citep{LIGORates}. The right axes show the percentage out of all detected events. \\ (A color version of this figure is available in the online journal.)}
\end{figure*}

Of $\sim$100,000 simulated sources spread across the 2015 and 2016 scenarios, $\approx1000$ events survived as confident GW detections.\footnote{There were slightly fewer surviving events in the 2016 configuration than in the 2015 configuration. This is because adding a third detector required us to apportion the two months of Gaussian noise to different combinations of detectors. In the 2015 simulation, all two months of data were allocated to the HL network. In 2016 about 43 days were devoted to the HLV and HL configurations, with the remaining 17 days of HV and LV mode contributing few detections.} No false alarms due to chance noise excursions survived our detection threshold; all events which should have been detectable were detected. We constructed probability sky maps using \textsc{bayestar} for all events and using \textsc{lalinference\_nest}/\textsc{mcmc} for a randomly selected subsample of 250 events from each scenario. Results from \textsc{lalinference\_bambi} are not shown because this sampler was run for only 30 events, and the sampling error bars would overwhelm the plots.\footnote{The three stochastic samplers \textsc{lalinference\_nest}/\textsc{mcmc}/\textsc{bambi} were interchangeable to the extent that they used the same likelihood and produced sky maps that agreed with each other.} The top four panels (a, b, c, d) of Figure~\ref{fig:area_hist} show cumulative histograms of the areas in deg$^2$ inside of the smallest 50\% and 90\% confidence regions for each event, for both sky localization methods. These contours were constructed using a `water\nobreakdashes-filling' algorithm: we sampled the sky maps using equal\nobreakdashes-area HEALPix \citep[Hierarchical Equal Area isoLatitude Pixelization;][]{HEALPix} pixels, ranked the pixels from most probable to least, and finally counted how many pixels summed to a given total probability. In the bottom two panels (e), (f) of Figure~\ref{fig:area_hist}, we also show a histogram of the smallest such constructed region that happened to contain the true location of each simulated source. We call this the searched area.

Panels (a--d) and (e), (f) may be thought of as measuring precision and accuracy respectively. The former measure how dispersed or concentrated each individual sky map is, while the latter describe how far the localization is from the true sky position. The 90\% area histograms and the searched area histograms also answer different but complementary questions that relate to two different strategies for following up LIGO/Virgo events. One might decide in 2015 to search for optical counterparts of all GW events whose 90\% areas are smaller than, for example, 200\,deg$^2$. By finding 200\,deg$^2$ on the horizontal axis of 90\% area histogram, one would find that this corresponds to following up $10\%$ of all GW detections. On the other hand, one might decide to always search the most probable 200\,deg$^2$ area for every GW event, corresponding to a different confidence level for every event. In this case, one would find 200\,deg$^2$ on the horizontal axis of the searched area histogram, and find that this strategy would enclose the true location of the GW source $64\%$ of the time.\footnote{One might naively expect that self\nobreakdashes-consistency would require the 90\% confidence area and searched area histograms to intersect at 90\% of detections, but this is not generally required because the posteriors of different events have widely different dimensions. However, it is true that 90\% of sources should be found within their respective 90\% confidence contours. This can be formalized into a graphical self\nobreakdashes-consistency test; see \citet{SiderySkyLocalizationComparison} for an example of application to GW parameter estimation.}

The left\nobreakdashes-hand axes of all four panels of Figure~\ref{fig:area_hist} show the expected cumulative number of detections, assuming the `realistic' BNS merger rates from~\citet{LIGORates}. We stress that the absolute detection rate might be two orders of magnitude smaller or one order of magnitude higher due to the large systematic uncertainty in the volumetric rate of BNS mergers, estimated from population synthesis and the small sample of Galactic binary pulsars~\citep{LIGORates}. An additional source of uncertainty in the detection rates is the Advanced LIGO/Virgo commissioning schedule given in \citet{LIGOObservingScenarios}. The proposed sensitivity in the 2016 scenario may be considered a plausible upper bound on the performance of the GW detector network in 2015, if commissioning occurs faster than anticipated. Likewise, the quoted sensitivity in the 2015 scenario is a plausible lower bound on the performance in 2016. The right\nobreakdashes-hand axes show the cumulative percentage of all detected sources. These percentages depend only on the gross features of the detector configuration and not on the astrophysical rates, so are relatively immune to the systematics described above.

Table~\ref{table:summary} summarizes these results.

\begin{deluxetable}{rr|rr|rr}
    \tablecaption{\label{table:summary}Summary of the 2015 and 2016 Scenarios, Listing the Participating Detectors, BNS Horizon Distance, Run Duration, and Fractions of Events Localized within 5, 20, 100, 200, or 500\,deg$^2$.}
    \tablecomments{A dash (---) represents less than 1\% of detections.}
    \tablehead{\colhead{} & \colhead{} & \multicolumn{2}{c}{2015} & \multicolumn{2}{c}{2016}}
    \startdata
    \multicolumn{2}{r}{Detectors} & \multicolumn{2}{c}{HL} & \multicolumn{2}{c}{HLV} \\
    \multicolumn{2}{r}{LIGO (HL) BNS range} & \multicolumn{2}{c}{54 Mpc} & \multicolumn{2}{c}{108 Mpc} \\
    \multicolumn{2}{r}{Run duration} & \multicolumn{2}{c}{3 months} & \multicolumn{2}{c}{6 months} \\
    \multicolumn{2}{r}{No. detections} & \multicolumn{2}{c}{0.091} & \multicolumn{2}{c}{1.5} \\
    \tableline
    \colhead{} & \colhead{} & \colhead{rapid} & \colhead{full PE} & \colhead{rapid} & \colhead{full PE} \\
    \tableline
    
\multirow{5}{*}{\parbox{1.4cm}{\raggedleft Fraction 50\% CR Smaller than}}
& 5\,deg$^2$&---&---&9\%&14\% \\
& 20\,deg$^2$&2\%&3\%&15\%&35\% \\
& 100\,deg$^2$&30\%&37\%&32\%&72\% \\
& 200\,deg$^2$&74\%&80\%&62\%&90\% \\
& 500\,deg$^2$&100\%&100\%&100\%&100\% \\
\tableline
\multirow{5}{*}{\parbox{1.4cm}{\raggedleft Fraction 90\% CR Smaller than}}
& 5\,deg$^2$&---&---&2\%&2\% \\
& 20\,deg$^2$&---&---&8\%&14\% \\
& 100\,deg$^2$&3\%&4\%&15\%&31\% \\
& 200\,deg$^2$&10\%&13\%&19\%&45\% \\
& 500\,deg$^2$&44\%&48\%&39\%&71\% \\
\tableline
\multirow{5}{*}{\parbox{1.4cm}{\raggedleft Fraction Searched Area Smaller than}}
& 5\,deg$^2$&3\%&4\%&11\%&20\% \\
& 20\,deg$^2$&14\%&19\%&23\%&44\% \\
& 100\,deg$^2$&45\%&54\%&47\%&71\% \\
& 200\,deg$^2$&64\%&70\%&62\%&81\% \\
& 500\,deg$^2$&87\%&89\%&83\%&93\% \\
\tableline
\multirow{3}{*}{Median Area $\bigg\{$}
& 50\% CR& 138\,deg$^2$& 124\,deg$^2$& 162\,deg$^2$& 43\,deg$^2$ \\
& 90\% CR& 545\,deg$^2$& 529\,deg$^2$& 621\,deg$^2$& 235\,deg$^2$ \\
& searched& 123\,deg$^2$& 88\,deg$^2$& 118\,deg$^2$& 29\,deg$^2$

    \enddata
\end{deluxetable}

\subsection{2015}
\label{sec:2015}

Our 2015 scenario assumes two detectors (HL) operating at an anticipated range of 54\,Mpc. About 0.1 detectable BNS mergers are expected, though there are nearly two orders of magnitude systematic uncertainty in this number due to the uncertain astrophysical rates. A detection in 2015 is possible, but likely only if the BNS merger rates or the detectors' sensitivity are on the modestly optimistic side. A typical or median event (with a localization area in the 50th percentile of all detectable events) would be localized to a 90\% confidence area of $\sim 500$\,deg$^2$.

We find that the area histograms arising from the \textsc{bayestar} rapid sky localization and the full parameter estimation agree within sampling errors, and that the sky maps resulting from the two analyses are comparable for any individual event. Put differently, the rapid sky localization contains nearly all of the information about sky localization for these events, with the full probability distributions over masses and spins becoming available when the stochastic samplers finish on a timescale of a day.

\begin{figure*}
    \begin{minipage}[b]{3.5in}
        \begin{center}
            \includegraphics{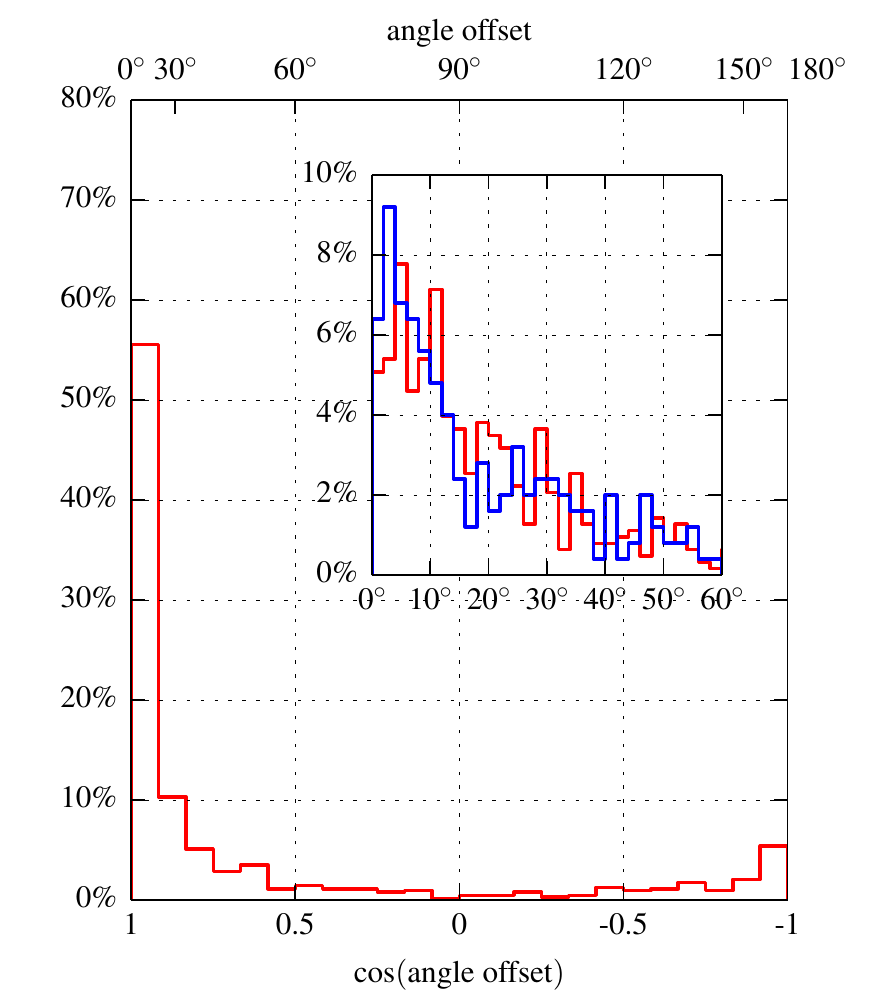}

            (a) 2015, HL
        \end{center}
    \end{minipage}
    \begin{minipage}[b]{3.5in}
        \begin{center}
            \includegraphics{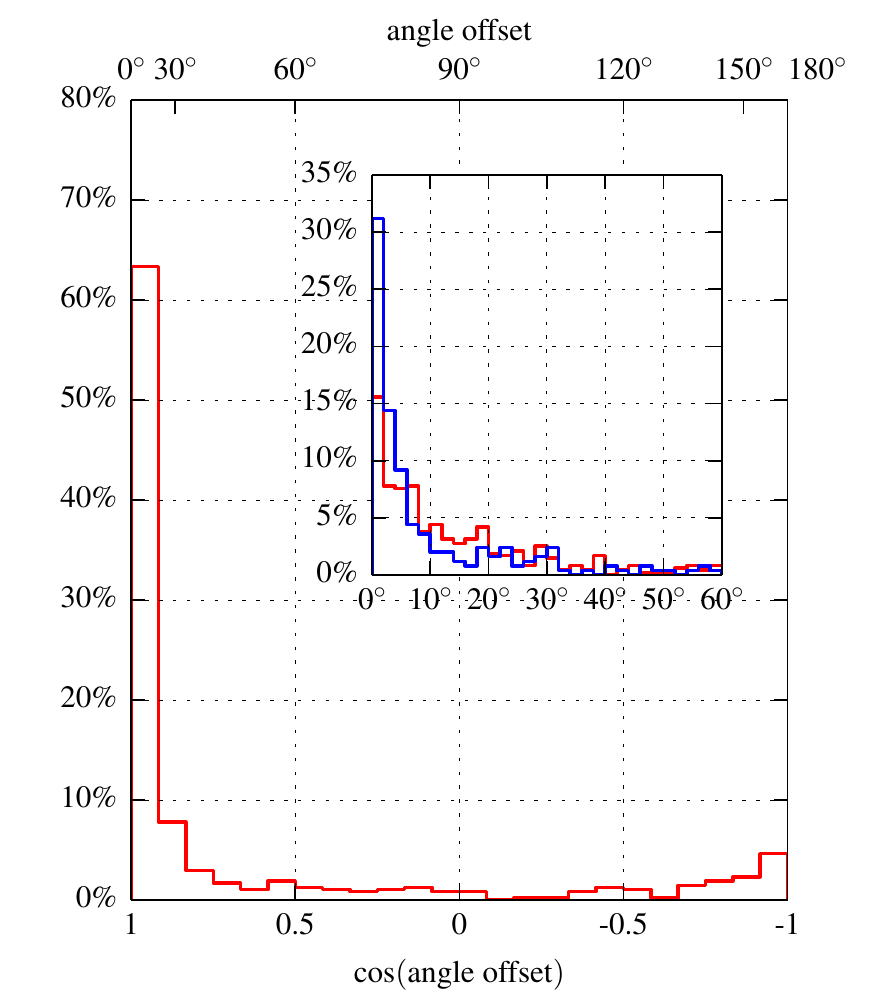}

            (b) 2016, HLV
        \end{center}
    \end{minipage}
    \caption{\label{fig:offset_hist}Normalized histogram of the cosine angular separation between the location of the simulated GW source and the \underline{maximum} a posteriori location estimate, for (a) the 2015 configuration and (b) the 2016 configuration. The red line encompasses all detections and their \textsc{bayestar} localizations, and the blue line the subsample of 250 events analyzed by \textsc{lalinference\_nest}/\textsc{mcmc}. The inset shows the distribution of angle offsets for angles less than 60$^\circ$. \\ (A color version of this figure is available in the online journal.)}
\end{figure*}

Figure~\ref{fig:offset_hist}(a) shows a histogram of the cosine of the angular separation between the true location of the simulated GW source and the maximum a posteriori estimate (the mode of the sky map, or the most probable location). The main feature is a peak at low separation. However, there is a second peak at the polar opposite of the true location, 180$^\circ$ away; about 15\% of events are recovered between 100 and 180$^\circ$ away from the true location.

Correspondingly, for any one event, it is common to find the probability distributed across two antipodal islands on opposite sides of the mean detector plane. We define this plane by finding the average of the two vectors that are normal to the planes of the two detectors' arms, and then taking the plane to which this vector is normal. This plane partitions the sky into two hemispheres. We find that one hemisphere is favored over the other by less than 20\% (which is to say that the odds favoring one hemisphere over the other are as even as 60\%/40\%) for 20\% of events.

The second peak admits a simple explanation as an unavoidable degeneracy due to the relative positions of the H and L interferometers. Before the Hanford and Livingston sites were selected, it was decided that the detectors' arms would be as closely aligned as possible~\citep[section V\nobreakdashes-C\nobreakdashes-2]{LIGOProposal}. Significant misalignment would have created patches of the sky that were accessible to one detector but in a null of the other detector's antenna pattern, useless for a coincidence test.

\begin{figure*}
    \caption{\label{fig:degeneracy}HL degeneracy. This, like all sky plots in this paper, is a Mollweide projection in geographic coordinates to emphasize spatial relationships with respect to the Earth\nobreakdashes-fixed GW detector network as well as possible ground\nobreakdashes-based telescope sites. Pluses denote the locations of signals whose best-estimate locations are offset by $\geq 100 ^\circ$, comprising the large\nobreakdashes-offset peak that is evident in Figure~\ref{fig:offset_hist}(a). The locations of zero time delay (simultaneous arrival at the H and L detectors) is shown as a thick black line. Shading indicates the \ac{RMS} network antenna pattern, with darker areas corresponding to high sensitivity and white corresponding to null sensitivity. \\ (A color version of this figure is available in the online journal.)}
    \includegraphics{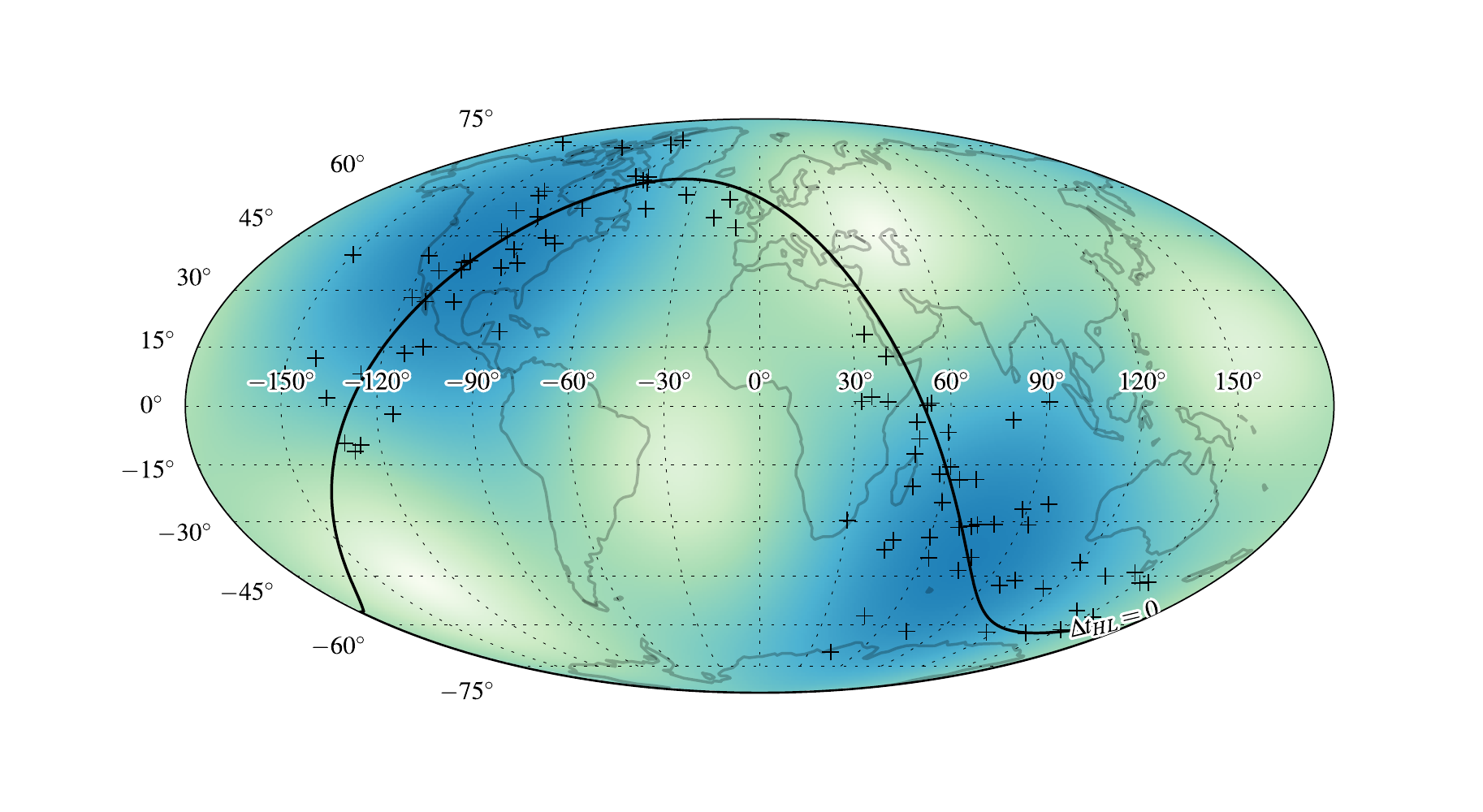}
\end{figure*}

The near alignment maximized the range of the detectors in coincidence, though at certain expenses for parameter estimation. Observe that the sensitivity of an interferometric GW detector is identical at antipodal points (i.e., symmetric under all rotations by 180$\arcdeg$). Therefore, any source that lies in the plane of zero time delay between the detectors is always localized to two opposite patches. Because the HL detectors were placed nearby (at continental rather than intercontinental distances) on the surface of the Earth so as to keep their arms nearly coplanar, their combined network antenna pattern has two maxima that lie on opposite sides of that great circle. As a consequence, a large fraction of sources are localized to two islands of probability that cannot be distinguished based on time or amplitude on arrival. See Figure~\ref{fig:degeneracy} for an illustration of these two degenerate patches.

A second undesirable side effect of the aligned antenna patterns is that GW polarization, observed as the phase difference on arrival at these two detectors, is of limited help for parameter estimation.

\begin{figure*}
    \caption{\label{fig:typical}Localization of a typical circa 2015 GW detection. This is a Mollweide projection in geographic coordinates. Shading is proportional to posterior probability per deg$^2$. This is a moderately loud event with $\rho_\mathrm{net}=15.0$, but its 90\% confidence area of 630\,deg$^2$ is fairly typical, in the 60th percentile of all detections. The sky map is bimodal with two long, thin islands of probability over the north and southern antenna pattern maxima. Neither mode is strongly favored over the other. Each island is forked like a snake's tongue, with one fork corresponding to the binary having face\nobreakdashes-on inclination ($\iota \approx 0^\circ$) and the other fork corresponding to face\nobreakdashes-off ($\iota \approx 180^\circ$). \\ This is event ID 18951 from Tables~\ref{table:2015-sim}~and~\ref{table:2015-coinc} and the online material (see the Appendix for more details). \\ (A color version of this figure is available in the online journal.)}
    \includegraphics{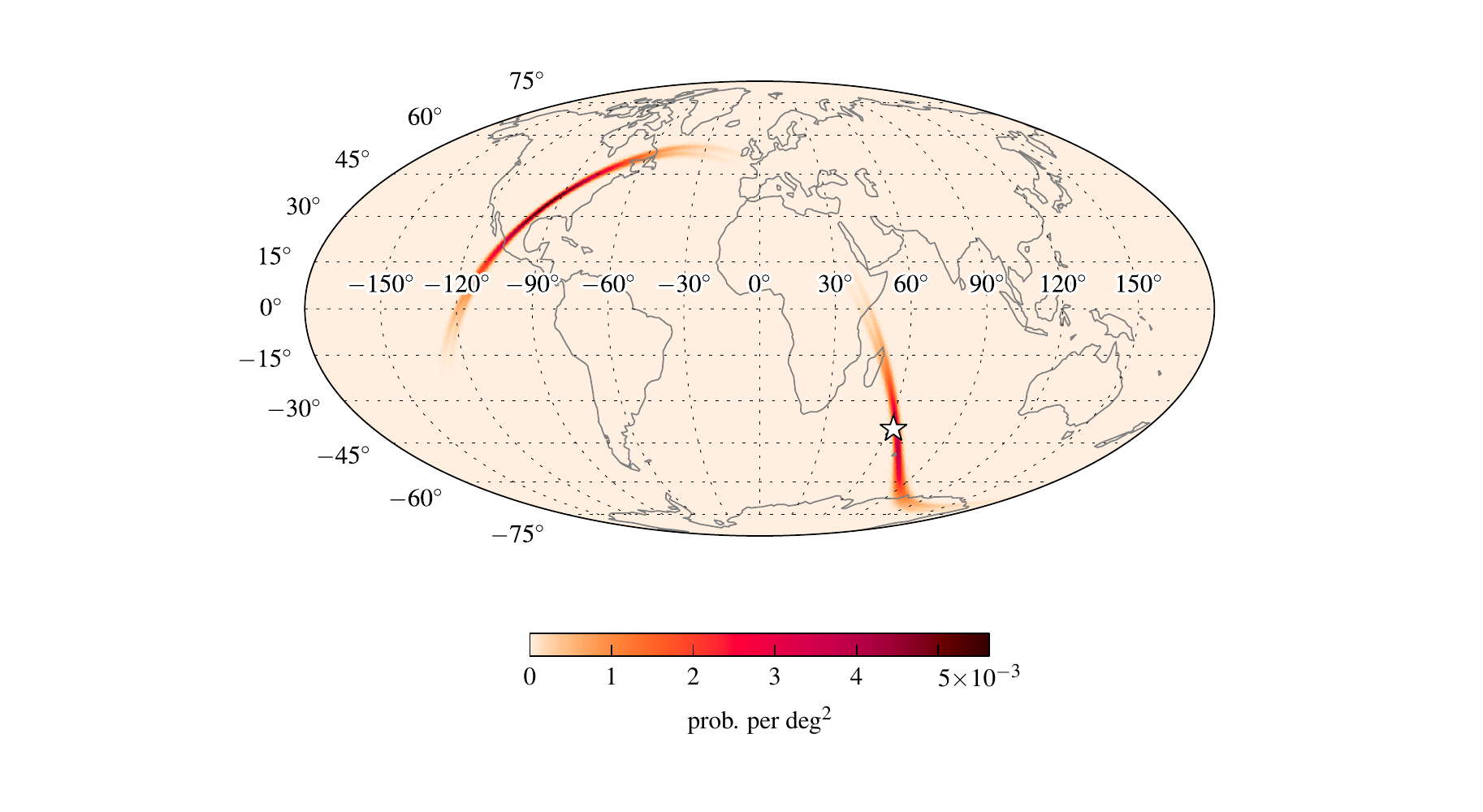}
\end{figure*}

A fairly typical sky map morphology, even at modestly high $\mathrm{SNR}_\mathrm{net}$, will consist of two extended arc-shaped modes, one over North America and a mirror image on the opposite side of the Earth. See Figure~\ref{fig:typical} for a typical event exhibiting this degeneracy. In this example, it is also possible to discern two narrow stripes resembling the forked tongue of a snake. This is a reflection of the HL network's limited polarization sensitivity. It occurs when the GW phases on arrival support two different binary inclination angles, with the orbital plane nearly facing the observer but with opposite handedness (usually peaked at $\iota \approx 30\arcdeg$ and $\iota \approx 150\arcdeg$; see \citealt{ShutzThreeFiguresOfMerit}). The two forks cross at a sky location where the GW data cannot distinguish between a clockwise or counterclockwise orbit.

The HL degeneracy is even apparent in earlier works on localization of GW bursts with networks of four or more detectors: \citet{CWBLocalization} drew a connection between accurate position reconstruction and sensitivity to both the `$+$' and `$\times$' GW polarizations, and noted that the close alignment of the HL detector network adversely affects position reconstruction. (They did not, however, point out the common occurrence of nearly $180^\circ$ errors, or note that the worst GW localizations paradoxically occur where the HL network's sensitivity is the greatest.)

\begin{figure*}
    \caption{\label{fig:typical-unimodal}Localization of a typical circa 2015 GW detection. This is a Mollweide projection in geographic coordinates. Shading is proportional to posterior probability per deg$^2$. This event's $\rho_\mathrm{net}=12.7$ is near the threshold, but its 90\% confidence area of 530\,deg$^2$ near the median. The sky map consists of a single, long, thin island exhibiting the forked-tongue morphology. \\ This is event ID 20342 from Tables~\ref{table:2015-sim}~and~\ref{table:2015-coinc} and the online material (see the Appendix for more details). \\ (A color version of this figure is available in the online journal.)}
    \includegraphics{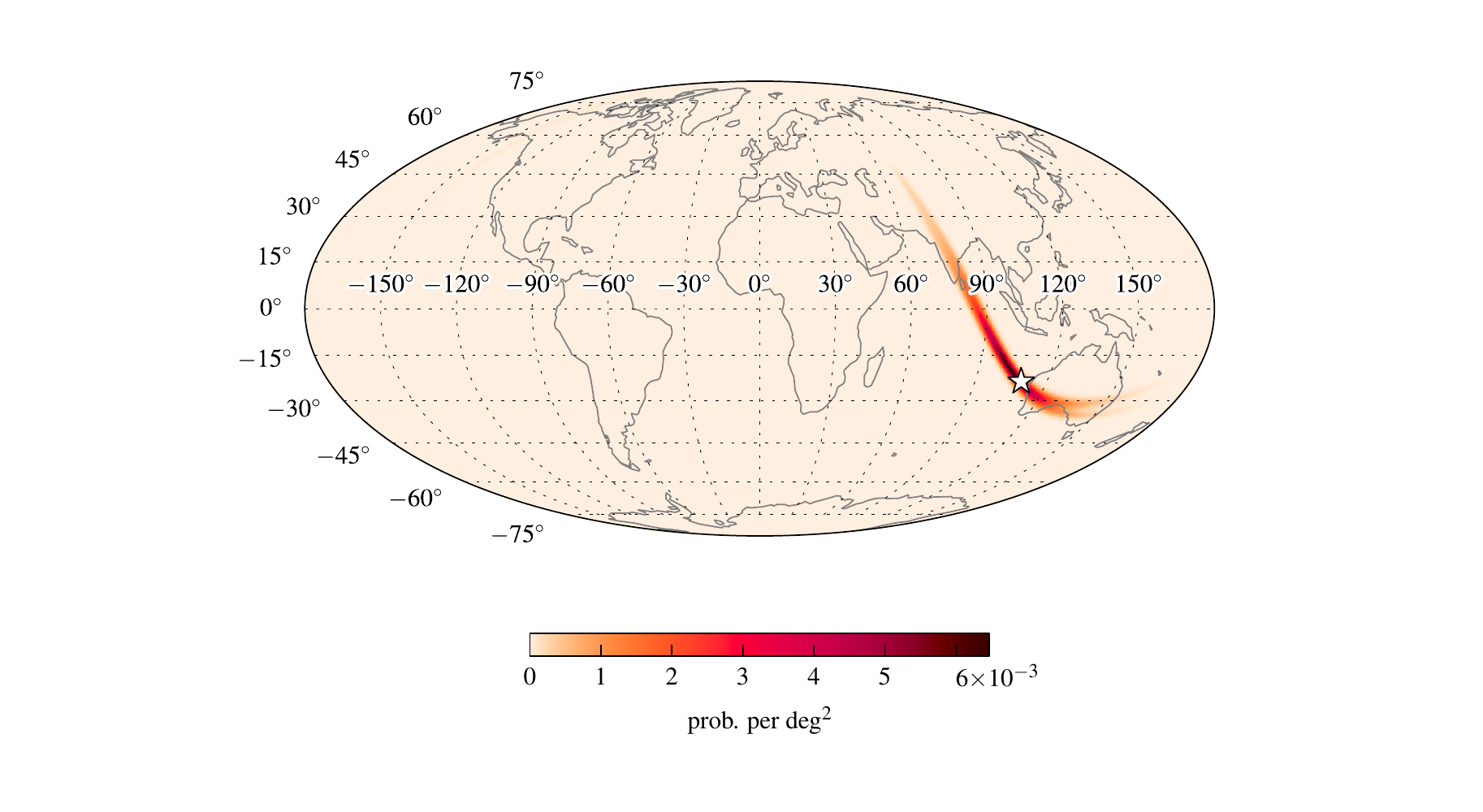}
\end{figure*}

The HL degeneracy affects most events that occur $\lesssim 30\arcdeg$ from one of the antenna pattern maxima. Most events that are $\gtrsim 50\arcdeg$ away have localizations that consist of a single long, thin arc or ring. See Figure~\ref{fig:typical-unimodal} for an example.

\begin{figure}
    \includegraphics{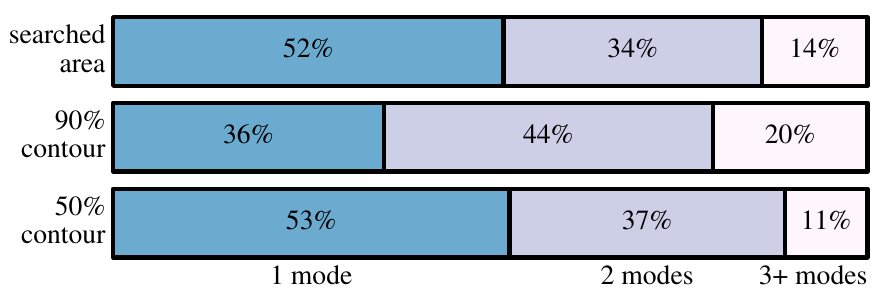}
    \caption{\label{fig:mode_hist}Frequency with which GW sky maps have one, two, or more disconnected modes during 2015. From top to bottom are the number of modes contained within the smallest confidence contour containing each simulated signal, the smallest 90\% contour, and the smallest 50\% contour. In 2015, roughly half of the sky maps will be unimodal, with most of the remainder being bimodal. \\ (A color version of this figure is available in the online journal.)}
\end{figure}

In Figure~\ref{fig:mode_hist}, we have plotted a histogram of the number of disconnected modes comprising the 50\% and 90\% confidence regions and the searched area, for the rapid localizations in the 2015 configuration. The ratios of events having one, two, or three or more modes depend weakly on the selected confidence level. In 2015, using either the 50\% contour or the searched area, we find that about half of events are unimodal and about a third are bimodal, the rest comprising three or more modes. Using the 90\% contour, we find that about a third of the events are unimodal and about half are bimodal.

\subsection{2016}

\begin{figure}
    \includegraphics{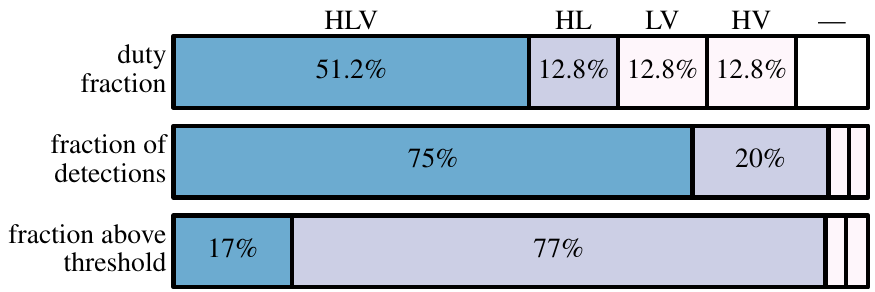}
    \caption{\label{fig:demographics}Breakdown of 2016 scenario by detector network. The top row shows the duty fraction of each subset of the detector network, the fraction of time when all three detectors~(HLV) are observing, when any pair of detectors are observing~(HL, LV, or HV), or when zero or one detector is observing~(---). The second row shows the fraction of coincident detections that occur under any given network~(HLV, HL, LV, or HV). The last row shows the fraction of coincident detections for which the given detectors have signals above the single-detector threshold of $\rho=4$.\\ (A color version of this figure is available in the online journal.)}
\end{figure}

In our 2016 scenario, the HL detectors double in range to 108\,Mpc and the V detector begins observations with a range of 36\,Mpc. Over this six\nobreakdashes-month science run we expect $\sim$1.5 detections, assuming a BNS merger rate of 1\,Mpc$^{-3}$\,Myr$^{-1}$. Figure~\ref{fig:demographics} shows how livetime and duty cycle breaks down according to detector network (HLV, HL, LV, or HV). About half of the time all three detectors are online, with the remaining time divided in four almost equal parts among the three pairs of detectors or $\leq 1$~detector. However, the HLV network accounts for about three\nobreakdashes-quarters of detections and the HL network for most of the rest.

When all three detectors (HLV) are operating, most detections are comprised of H and L triggers, lacking a trigger from V because the signal is below the single\nobreakdashes-detector threshold of $\rho=4$. Slightly more than half (57\%) of all detections have a signal below threshold in one operating detector (almost always~V) while slightly less than half (43\%) consist of triggers from all operating detectors.

The first half consists mainly of HLV events that are detected by HL but not Virgo. For these events, the stochastic samplers provide a marked improvement in sky localization; their 90\% confidence regions have about one\nobreakdashes-third as much area as their rapid localizations. This is because the rapid localization makes use of only the triggers provided by the detection pipeline, lacking information about the signal in Virgo if its \ac{SNR} is $< 4$. The stochastic samplers, on the other hand, can use data from all operating detectors, regardless of \ac{SNR}. Therefore, in the present analysis, an improved sky localization would be available for half of the detections on a timescale of a day. Fortunately, for BNS sources, it is immediately known whether an improved localization is possible, since this statement only depends on what detectors were online and which contributed triggers. On the other hand, it may be possible to provide prompt sky localizations for all events by simply lowering the single\nobreakdashes-detector threshold. If the single-detector threshold was dropped to unity, essentially no event would lack a Virgo trigger. There are also efforts to do away with the single\nobreakdashes-detector threshold entirely \citep{2012PhRvD..86l3010K,2013arXiv1307.4158K}. Simultaneously, there is promising work under way in speeding up the full parameter estimation using reduced order quadratures \citep{roq-pe}, interpolation \citep{interpolation-pe}, jump proposals that harness knowledge of the multimodal structure of the posterior \citep{KDEJumpProposal}, hierarchical techniques \citep{HierarchicalParameterEstimation}, and machine learning techniques to accelerate likelihood evaluation \citep{BAMBI,SKYNET}. It seems possible that the the delayed improvement in sky localization may be a temporary limitation that can be overcome well before 2016.

The second half consists of HLV events with triggers from all three detectors and events that occur when only HL, HV, or LV are operating. For these, the \textsc{bayestar} analysis and the full stochastic samplers produce comparable sky maps.

\begin{figure*}
    \caption{\label{fig:typical-hlv}Localization of a typical circa 2016 GW detection in the HLV network configuration. This is a Mollweide projection in geographic coordinates. This event consists of triggers in H and L and has $\rho_\mathrm{net} = 13.4$. The rapid sky localization gives a 90\% confidence region with an area of 1100\,deg$^2$ and the full stochastic sampler gives 515\,deg$^2$. \\ This is event ID 821759 from Tables~\ref{table:2016-sim}~and~\ref{table:2016-coinc} and the online material (see the Appendix for more details). \\ (A color version of this figure is available in the online journal.)}
    \begin{center}
        \includegraphics{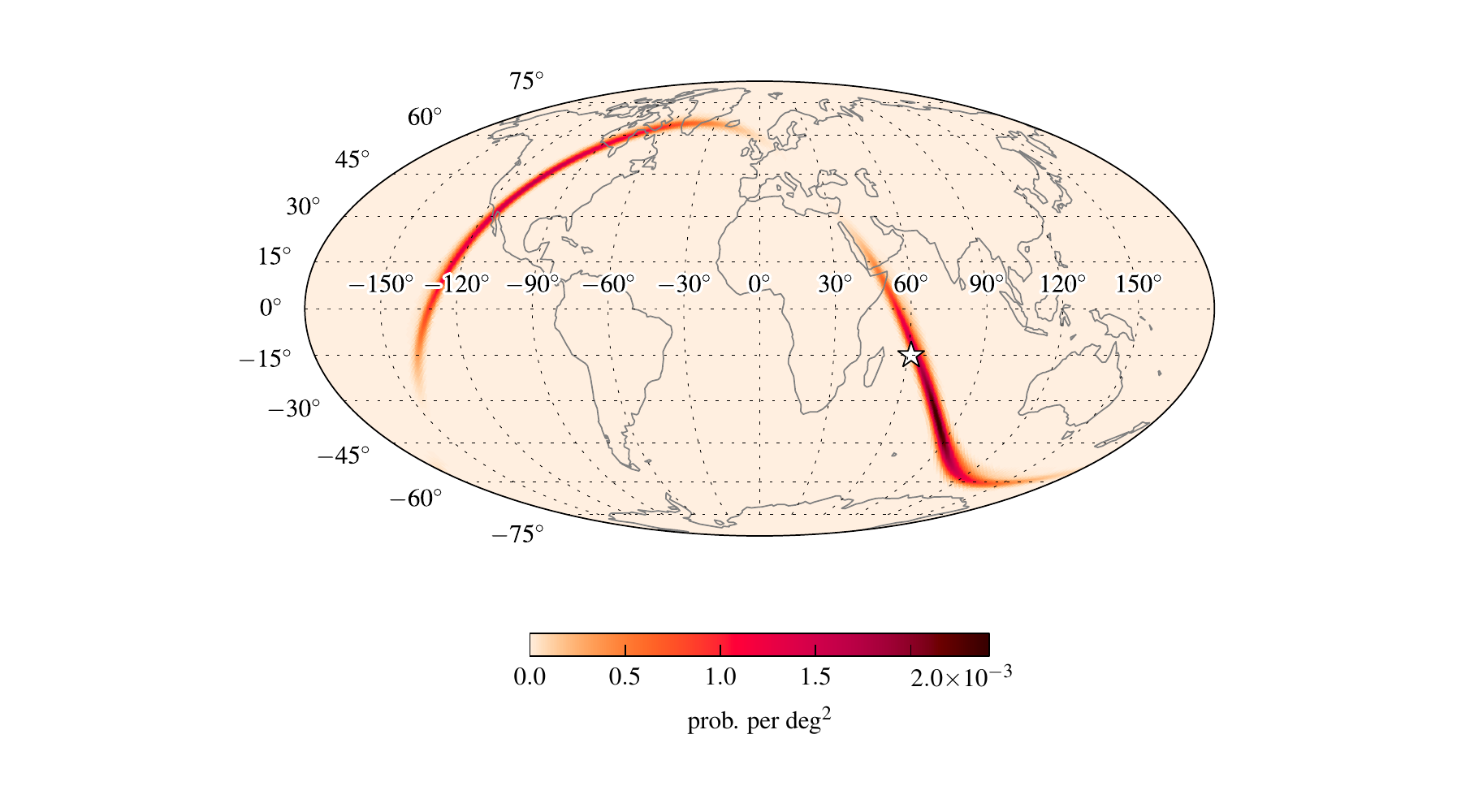}

        (a) \textsc{bayestar}

        \includegraphics{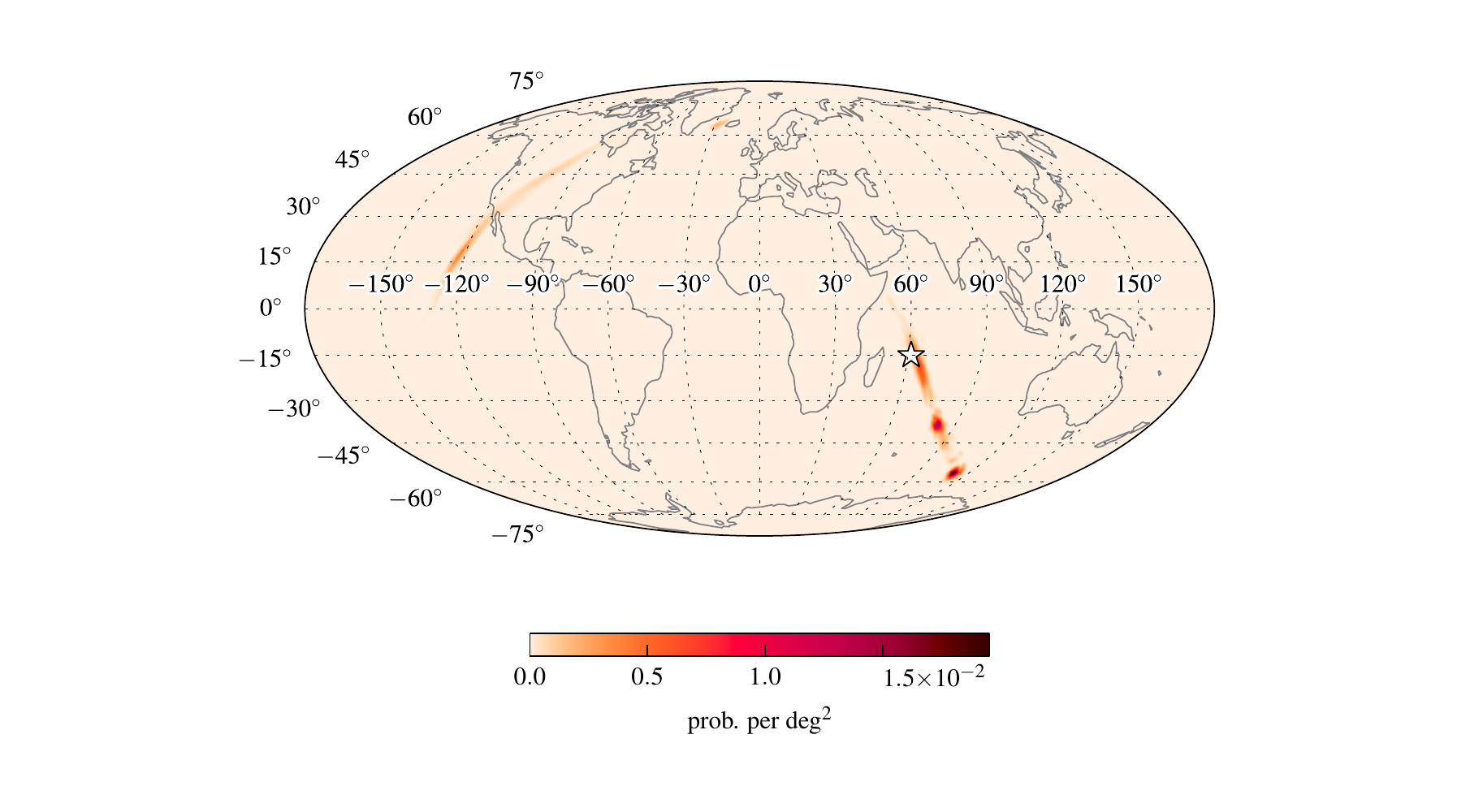}

        (b) \textsc{lalinference\_mcmc}
    \end{center}
\end{figure*}

For nearby loud sources ($\rho_\mathrm{net} \gtrsim 20$), the HLV network frequently produces compact sky localizations concentrated in a single island of probability. However at low SNR ($\rho_\mathrm{net} \lesssim 20$), and especially for the events that are detected as only double coincidence (HL), the refined localizations from the full stochastic samplers often identify many smaller modes. An $\rho_\mathrm{net} =13.4$ example is shown in Figure~\ref{fig:typical-hlv}. In this event, the rapid sky localization has two modes and has a morphology that is well\nobreakdashes-described by the HL degeneracy explained in Section~\ref{sec:2015}. However, the refined, full parameter estimation breaks this into at least four smaller modes.

Of the remaining events, most occur when only the two HL detectors are operating. These look qualitatively the same as the 2015 case; their sky maps generally exhibit one or two slender islands of probability. However, percentage\nobreakdashes-wise, two\nobreakdashes-detector events are localized worse in the 2016 scenario than in the 2015 scenario. This unusual result is easily explained. Though the LIGO detectors improve in sensitivity at every frequency, with the particular noise curves that we assumed, the signal bandwidth is actually slightly lower with the 2016 sensitivity compared to 2015. This is because of improved sensitivity at low frequency. Applying Equation~(\ref{eq:sigmat}), we find that for a $(1.4, 1.4)\,M_\odot$ binary at $\rho=10$, one of the 2015 LIGO detectors has an \ac{RMS} timing uncertainty of 131\,{\textmu}s, whereas one of the 2016 LIGO detectors has a timing uncertainty of 158\,{\textmu}s. Clearly, the 2016 detectors will produce more constraining parameter estimates for sources at any fixed distance as the \ac{SNR} improves. However, for constant \ac{SNR} the 2016 LIGO detectors should find areas that are $(158/131)^2=1.45$ times larger than events at the same \ac{SNR} in 2015. This is indeed what we find.

\begin{figure*}
    \caption{\label{fig:typical-hv}Rapid localization of a typical circa 2016 GW detection in the HV network configuration. This is a Mollweide projection in geographic coordinates. This event's $\rho_\mathrm{net} = 12.2$ is near the detection threshold. Its 90\% confidence area is 4600\,deg$^2$, but the true position of the source (marked with the white pentagram) is found after searching 65\,deg$^2$. \\ This is event ID 655803 from Tables~\ref{table:2016-sim}~and~\ref{table:2016-coinc} and the online material (see the Appendix for more details). \\ (A color version of this figure is available in the online journal.)}
    \includegraphics{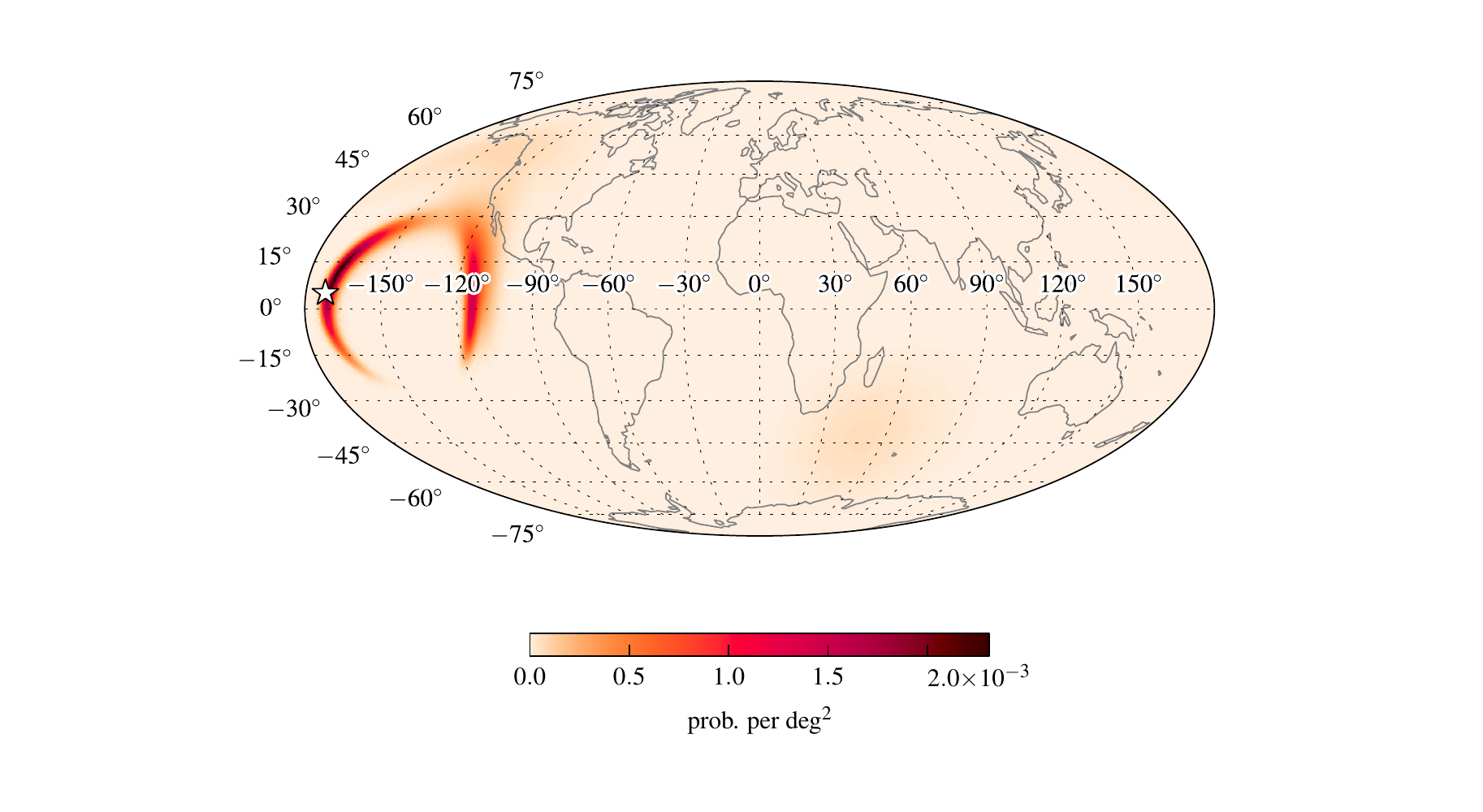}
\end{figure*}

Two\nobreakdashes-detector events involving Virgo (HV and LV) are rare, accounting for only about 6\% of detections. Sky maps for these events sometimes exhibit multiple fringes spread over a quadrant of the sky. These are in part due to the increased importance of phase\nobreakdashes-on\nobreakdashes-arrival due to the oblique alignment of the LIGO and Virgo antenna patterns, which gives the network a limited ability to measure GW polarization. Occasionally there are also diffuse clouds of probability near the participating LIGO detector's two antenna pattern maxima, which may be a vestige of the antenna pattern. A typical HV event that exhibits both features is shown in Figure~\ref{fig:typical-hv}.

\section{Discussion}

\subsection{Caveats}

We reiterate that the scenarios we have described make assumptions about the astrophysical rate of BNS mergers and the Advanced LIGO/Virgo sensitivity as a function of time. The former is subject to orders of magnitude uncertainty due to the small sample of known galactic binary pulsars as well as model dependence in population synthesis~\citep{LIGORates}. The latter could deviate from \citet{LIGOObservingScenarios} depending on actual Advanced LIGO/Virgo commissioning progress. However, the fractions of events localized within a given area are robust with respect to both of these effects.

We have dealt only with BNS mergers. NSBH mergers are also promising sources for closely related GW signals and EM transients. A similar, comprehensive investigation of GW sky localization accuracy for NSBH signals is warranted.

In this simulation, we have used ideal Gaussian noise, but selected a detection threshold that is designed to reproduce expected performance in detectors with realistically wide tails due to instrumental and environmental glitches. If Advanced LIGO's and Virgo's improved seismic isolation and control systems are even more effective at suppressing such glitches than their initial counterparts were, then the $\rho_\mathrm{net}$ threshold for confident detection would decrease, yielding discoveries earlier but with larger typical sky localization areas.

We remind the reader that the events comprising this study would be regarded as confident detections, with $\text{FAR} \lesssim 10^{-2}$\,yr$^{-1}$, based on GW observations alone. In practice, some observers may choose to follow up more marginal detection candidates. For instance, a group with enough resources and telescope time to follow up one candidate per month might filter events with $\textsc{FAR} \leq 12$\,yr$^{-1}$. A high false alarm rate threshold will admit correspondingly lower $\rho_\mathrm{net}$ candidates with coarser localizations than what we have presented here.

Finally, on a positive note, the number of detections is expected to increase considerably as commissioning proceeds toward final design sensitivity. Furthermore, sky localization will improve radically as the HLV detectors approach comparable sensitivity. The addition of two more planned ground\nobreakdashes-based GW detectors, LIGO\nobreakdashes--India and KAGRA, would likewise increase rates and improve sky localizations dramatically \citep{ShutzThreeFiguresOfMerit,Veitch:2012,FairhurstLIGOIndia,NissankeKasliwalEMCounterparts,LIGOObservingScenarios}.

\subsection{Detection Scenarios}

From our representative sample of hundreds of early Advanced LIGO/Virgo events emerge a few common morphologies and several possible scenarios for the early detections of GWs from a BNS merger.

We find that in both 2015 and 2016, the detection rate is highly anisotropic, proportional to the cube of the network antenna pattern with a strong excess above North America and the Indian Ocean and deficits in four spots over the south Pacific, south Atlantic, Caucasus, and north Pacific.

\begin{enumerate}

\item \label{item:hl-unimodal} \textit{HL event, single arc}---This scenario is relevant for the HL network configuration and applies to both 2015 and 2016. Figure~\ref{fig:typical-unimodal} shows a typical sky map for a near\nobreakdashes-threshold detection with $\rho_\mathrm{net} = 12.7$, exhibiting a single long, extended arc spanning $\sim$500\,deg$^2$.

\item\label{item:hl-bimodal} \textit{HL event, two degenerate arcs}---This scenario also applies to 2015 or to HL livetime in 2016. Figure~\ref{fig:typical} shows a typical sky map with a moderately high $\rho_\mathrm{net} = 15.0$, localized to $\sim$600\,deg$^2$. Its localization embodies the HL degeneracy, with two strong, long, thin modes over North America and the Indian Ocean, separated by nearly $180^\circ$ and therefore 12\,hr apart in hour angle. Inevitably, one of these two modes will be nearly Sun\nobreakdashes-facing and inaccessible to optical facilities. Because of the bimodality, these sky maps can span slightly larger areas than case~\ref{item:hl-unimodal}. After taking an inevitable 50\% hit in visibility, such events resemble the single arc scenario.

Whether a given source falls into scenario \ref{item:hl-unimodal} or \ref{item:hl-bimodal} is largely determined by its sky location relative to the network antenna pattern. The transition occurs between $\sim 30\arcdeg$ and $\sim 50\arcdeg$ away from the two points of maximum sensitivity.

\item \textit{HLV event, degeneracy broken by Virgo}---This scenario applies only to the 2016 configuration, while all three instruments are online. The rapid sky localization looks similar to the previous scenario, a pair of long, thin rings over the northern and southern hemispheres, but the full parameter estimation cuts this down to a handful of islands of probability covering as little as half to a third of the area, $\sim 200$\,deg$^2$. For such an event, the refined localization could be used to guide $\sim$day\nobreakdashes-cadence kilonova\nobreakdashes-hunting observations or to re\nobreakdashes-target the vetting of afterglow candidates arising from early\nobreakdashes-time observations. Several wide\nobreakdashes-field facilities could be employed to monitor modes that lie in different hemispheres.

\item \textit{HLV event, compactly localized}---This is another 2016, three-detector scenario. It describes many of the events that are detected with triggers in all three instruments. These are many of the best\nobreakdashes-localized events, with 90\% confidence regions only a few times 10\,deg$^2$ in area. These events are generally localized to one simply connected region and exhibit a less pronounced preference for particular sky locations. In this scenario, it is most likely that the rapid sky localization and the full parameter estimation will be similar. This is observationally the simplest scenario: just one of the several wide\nobreakdashes-field optical searches (for instance, ZTF or BlackGEM) would be able to scan the whole error region at a daily cadence.

\end{enumerate}

\subsection{Comparison with Other Studies}

This is the first study so far to combine an astrophysically motivated source population, realistic sensitivity and detector network configurations, event selection effects arising from a genuine detection pipeline instead of an ad hoc threshold, and parameter estimation algorithms that will be deployed for GW data analysis. This study also has a much larger sample size and lower statistical uncertainty than most of the prior work. It is therefore somewhat difficult to compare results to other studies which each have some but not all of these virtues.

To the best of the authors' knowledge, \citet{Raymond:2009} were the first to point out the power of Bayesian priors for breaking sky degeneracies in two\nobreakdashes-detector networks, challenging a prevailing assumption at the time that two detectors could only constrain the sky location of a compact binary signal to a degenerate annulus. \citet{LIGOObservingScenarios} speculated that two\nobreakdashes-detector, 2015, HL configuration sky maps would be rings spanning ``hundreds to thousands'' of deg$^2$, that coherence and amplitude consistency would ``sometimes'' resolve the localizations to shorter arcs. With our simulations, we would only revise that statement to read ``hundreds to \emph{a} thousand'' deg$^2$ and change ``sometimes'' to ``always.'' \citet{KasliwalTwoDetectors} recently argued for the feasibility of optical transient searches (in the context of kilonovae) with two\nobreakdashes-detector GW localizations.

\citet{LIGOObservingScenarios} used time\nobreakdashes-of\nobreakdashes-arrival triangulation \citep{FairhurstTriangulation} to estimate the fraction of sources with 90\% confidence regions smaller than 20\,deg$^2$, finding a range of 5\nobreakdashes--12\% for 2016. We find 14\%. Our values are more optimistic, but perhaps also more realistic for the assumed detector sensitivity. Our sky localization takes into account phase and amplitude information, which \citet{Grover:2013} points out can produce $\approx$0.4 times smaller areas compared to timing alone. However, it is clear from both \citet{LIGOObservingScenarios} and our study that such well\nobreakdashes-localized events will comprise an exceedingly small fraction of GW detections until the end of the decade. We therefore echo \citet{KasliwalTwoDetectors} in stressing the importance of preparing to deal with areas of hundreds of deg$^2$ in the early years of Advanced LIGO and Virgo.

\citet{NissankeKasliwalEMCounterparts} used an astrophysical distance distribution, drawing source positions uniformly in comoving volume for distances $d_\mathrm{L} > 200$\,Mpc and from a $B$\nobreakdashes-band luminosity\nobreakdashes-weighted galaxy catalog for distances $d_\mathrm{L} \leq 200$\,Mpc. They generated sky maps using their own MCMC code. Similar to this study, they imposed a threshold of $\rho_\mathrm{net} > 12$. They explore several different GW detector network configurations. The most similar to our 2016 scenario is an HLV network at final design sensitivity. They find a median 95\% confidence region area of $\sim$20\,deg$^2$. For comparison, we find a 95\,deg$^2$ confidence area of 374\,deg$^2$. Our much larger number is explained by several factors. First, we did not draw nearby sources from a galaxy catalog, so we have fewer loud, nearby sources. Second, since we accounted for duty cycle, poorly localized two\nobreakdashes-detector events account for a quarter of our sample. Third, and most important, we assumed Advanced Virgo's anticipated initial sensitivity rather than its final design sensitivity.

\citet{RodriguezBasicParameterEstimation} also studied an HLV network at final design sensitivity. Their simulated signals had identically zero noise content, the average noise contribution among all realizations of zero-mean Gaussian noise. All of their simulated events had a relatively high $\rho_\mathrm{net} = 20$. They find a median $95\%$ confidence area of 11.2\,deg$^2$. If we consider all of our 2016 scenario HLV events with $19.5 \leq \rho_\mathrm{net} \leq 20.5$, we find a median area of 126\,deg$^2$. Our significantly larger number is once again partly explained by our less sensitive Virgo detector, which introduces several multimodal events even at this high $\rho_\mathrm{net}$.

Similarly, \citet{Grover:2013} and \citet{SiderySkyLocalizationComparison} studied a three\nobreakdashes-detector network, but at initial LIGO design sensitivity. These studies were primarily concerned with evaluating Bayesian parameter estimation techniques with respect to triangulation methods. They found much smaller areas, with a median of about 3\,deg$^2$. Both papers used a source population that consisted mainly of very high\nobreakdashes-\ac{SNR} signals with binary black hole masses, with distances distributed logarithmically. All of these effects contribute to unrealistically small areas.

Finally, \citet{KasliwalTwoDetectors} made the first small\nobreakdashes-scale systematic study of localizability with two LIGO detectors, albeit at final Advanced LIGO design sensitivity. For these noise curves and a $(1.4, 1.4)\,M_\odot$ binary with single-detector $\rho=10$, Equation~(\ref{eq:sigmat}) gives a timing uncertainty of 142\,{\textmu}s. Their different choice of noise curves should result in areas that are $(131/142)^2 \approx 0.85$ times smaller than ours, at a given $\rho_\mathrm{net}$. As we have, they imposed a network \ac{SNR} threshold of $\rho_\mathrm{net} \geq 12$ on all detections\footnote{\citet{NissankeKasliwalEMCounterparts} and \citet{KasliwalTwoDetectors} present a parallel set of results for a threshold $\rho_\mathrm{net} > 8.5$, relevant for a coherent GW search described by \citet{harry-single-spin}. However, the coherent detection statistic described by \citet{harry-single-spin} is designed for targeted searches at a known sky location (for instance, in response to a GRB). Thus the $\rho_\mathrm{net} > 8.5$ threshold is not relevant for optical follow\nobreakdashes-up triggered by a detection from an all\nobreakdashes-sky GW search. Furthermore, this reduced threshold is not relevant to the HL configuration because the coherent detection statistic reduces to the network \ac{SNR} for networks of two detectors or fewer.}. They find a median 95\% confidence area of $\sim 250$\,deg$^2$ from a catalog of 17 events. From our 2015 scenario, we find a median 90\% confidence area that is almost twice as large, $\sim$500\,deg$^2$. Though we cannot directly compare our 90\% area to their 95\% area, our 95\% area would be even larger. Several factors could account for this difference, including the smaller sample size in \citet{KasliwalTwoDetectors}. Also \citet{KasliwalTwoDetectors}, like \citet{NissankeKasliwalEMCounterparts}, drew nearby sources from a galaxy catalog to account for clustering, so their population may contain more nearby, well\nobreakdashes-localized events than ours. Another difference is that \citet{KasliwalTwoDetectors} do not report any multimodal localizations or the 180$^\circ$ degeneracy that we described in Section~\ref{sec:2015}.

\subsection{Conclusion}

Many previous sky localization studies have found that networks of three or more advanced GW detectors will localize BNS mergers to tens of deg$^2$. 
However, given realistic commissioning schedules, areas of hundreds of deg$^2$ will be typical in the early years of Advanced LIGO and Virgo.

We caution that multimodality and long, extended arcs will be a common and persistent feature of Advanced LIGO/Virgo detections. We caution that existing robotic follow\nobreakdashes-up infrastructure designed for GRBs, whose localizations are typically nearly Gaussian and unimodal, will need to be adapted to cope with more complicated geometry. In particular, we advise optical facilities to evaluate the whole GW sky map when determining if and when a given event is visible from a particular site.

We have elucidated a degeneracy caused by the relative orientations of the two LIGO detectors, such that position reconstructions will often consist of two islands of probability separated by 180$^\circ$. We have shown that this degeneracy is largely broken by adding Virgo as a third detector, even with its significantly lower sensitivity. We have shown that sub\nobreakdashes-threshold GW observations are important for sky localization and parameter estimation.

We have demonstrated a real\nobreakdashes-time detection, sky localization, and parameter estimation infrastructure that is ready to deliver Advanced LIGO/Virgo science. The current analysis has some limitations for the three\nobreakdashes-detector network, an undesirable trade\nobreakdashes-off of sky localization accuracy and timescale. Work is ongoing to lift these limitations by providing the rapid sky localization with information below the present single\nobreakdashes-detector threshold and by speeding up the full parameter estimation by a variety of methods \citep{roq-pe,interpolation-pe,KDEJumpProposal,HierarchicalParameterEstimation,BAMBI}.

We have exhibited an approach that involves three tiers of analysis, which will likely map onto a sequence of three automated alerts with progressively more information on longer timescales, much as how observers in the GRB community are used to receiving a sequence of Gamma\nobreakdashes-ray Coordinates Network (GCN) notices about a high\nobreakdashes-energy event.

The maximum timescale of the online GW analysis, about a day, is appropriate for searching for kilonova emission. However, the availability of \textsc{bayestar}'s rapid localizations within minutes of a merger makes it possible to search for X\nobreakdashes-ray and optical emission. Due to jet collimation, these early\nobreakdashes-time signatures are expected to accompany only a small fraction of LIGO/Virgo events. However, the comparative brightness and distinctively short timescale of the optical afterglow makes it an attractive target. PTF has recently proved the practicality of wide\nobreakdashes-field afterglow searches through the blind discovery of afterglow\nobreakdashes-like optical transients \citep{PTF11agg,iPTF14yb} and the detection of optical afterglows of \emph{Fermi} GBM bursts \citep{iPTF13bxl}. We encourage optical transient experiments such as ZTF and BlackGEM to begin searching for EM counterparts promptly based on the rapid GW localization. In the most common situation of no afterglow detection, the early observations may be used as reference images for longer\nobreakdashes-cadence kilonova searches.

\acknowledgements A catalog of the sky maps used in this study is available from \url{http://www.ligo.org/scientists/first2years}. See the Appendix for more details.

L.P.S. and B.F. thank generous support from the National Science Foundation (NSF) in the form of Graduate Research Fellowships. B.F. acknowledges support through NSF grants DGE\nobreakdashes-0824162 and PHY\nobreakdashes-0969820. A.L.U. and C.P. gratefully acknowledge NSF support under grants PHY\nobreakdashes-0970074 and PHY\nobreakdashes-1307429 at the University of Wisconsin\nobreakdashes--Milwaukee (UWM) Research Growth Initiative. J.V. was supported by the research programme of the Foundation for Fundamental Research on Matter (FOM), which is partially supported by the Netherlands Organisation for Scientific Research (NWO), and by Science and Technology Facilities Council (STFC) grant ST/K005014/1. P.G. is supported by a NASA Postdoctoral Fellowship administered by the Oak Ridge Associated Universities.

\textsc{gstlal} analyses were produced on the NEMO computing cluster operated by the Center for Gravitation and Cosmology at UWM under NSF Grants PHY\nobreakdashes-0923409 and PHY\nobreakdashes-0600953. The \textsc{bayestar} analyses were performed on the LIGO\nobreakdashes--Caltech computing cluster. The MCMC computations were performed on Northwestern's CIERA High\nobreakdashes-Performance Computing cluster GRAIL\footnote{\url{http://ciera.northwestern.edu/Research/Grail\_Cluster.php}}.

We thank Patrick Brady, Vicky Kalogera, Erik Katsavounidis, Richard O'Shaughnessy, Ruslan Vaulin, and Alan Weinstein for helpful discussions.

This research made use of Astropy\footnote{\url{http://www.astropy.org}} \citep{astropy}, a community-developed core Python package for Astronomy. Some of the results in this paper have been derived using HEALPix \citep{HEALPix}. Public-domain cartographic data is courtesy of Natural Earth\footnote{\url{http://www.naturalearthdata.com}} and processed with MapShaper\footnote{\url{http://www.mapshaper.org}}.

\textsc{bayestar},
\textsc{lalinference\_nest}, and
\textsc{lalinference\_mcmc}
are part of the LIGO Algorithm Library Suite\footnote{\url{http://www.lsc-group.phys.uwm.edu/cgit/lalsuite/tree}} and the LIGO parameter estimation package, \textsc{LALInference}. Source code for
\textsc{gstlal}%
\footnote{\url{http://www.lsc-group.phys.uwm.edu/cgit/gstlal/tree/}}
and \textsc{LALInference}%
\footnote{\url{http://www.lsc-group.phys.uwm.edu/cgit/lalsuite/tree/lalinference}}
are available online under the terms of the GNU General Public License.

LIGO was constructed by the California Institute of Technology and Massachusetts Institute of Technology with funding from the NSF and operates under cooperative agreement PHY\nobreakdashes-0757058.

This is LIGO document number {LIGO}\nobreakdashes-{P1300187}\nobreakdashes-{v25}.

\appendix

We describe a catalog of all simulated events, detections, and sky maps that were generated in this study.

For the 2015 scenario, parameters of simulated signals are given in Table~\ref{table:2015-sim}. In the same order, parameters of the detection including the operating detector network, false alarm rate, $\rho_\mathrm{net}$, \ac{SNR} in each detector, recovered masses, and sky localization areas are given in Table~\ref{table:2015-coinc}. For the 2016 scenario, the simulated signals are recorded in Table~\ref{table:2016-sim} and the detections in Table~\ref{table:2016-coinc}. In the print journal, parameters are given for just the four sample events that appear earlier in the text (see Figures~\ref{fig:typical},~\ref{fig:typical-unimodal},~\ref{fig:typical-hlv},~and~\ref{fig:typical-hv}). In the machine readable tables in the online journal, parameters are given for all detected signals.

The tables give two integer IDs. The ``event ID'' column corresponds to a field in that scenario's full \textsc{gstlal} output that identifies one coincident detection candidate. The ``simulation ID'' likewise identifies one simulated signal. In the full \textsc{gstlal} output, there may be zero or many event candidates that match any given simulated signal. However, in our catalog there is one\nobreakdashes-to\nobreakdashes-one correspondence between simulation and event IDs because we have retained only simulated signals that are detected above threshold, and only the highest \ac{SNR} detection candidate for each signal.

Note that the arbitrary dates of the simulated signals range from August~21 through 2010~October~19. This reflects the two\nobreakdashes-month duration of the simulated data stream, not the dates or durations of the anticipated Advanced LIGO/Virgo observing runs.

For convenience, we also provide a browsable sky map catalog\footnote{\url{http://www.ligo.org/scientists/first2years}}. This Web page provides a searchable version of Tables~\ref{table:2015-sim}\nobreakdashes--\ref{table:2016-coinc}, with posterior sky map images from both the rapid parameter estimation and the stochastic samplers.

The Web page also provides, for each localization, a FITS file representing the posterior in the HEALPix projection~\citep{HEALPix} using the NESTED indexing scheme. For reading these files, the authors recommend the Python package Healpy\footnote{\url{http://healpy.readthedocs.org}} or the HEALPix C/C++/IDL/Java/Fortran library\footnote{\url{http://healpix.sourceforge.net}}. They can also be displayed by many standard imaging programs such as DS9\footnote{\url{http://ds9.si.edu}} and Aladin\footnote{\url{http://aladin.u-strasbg.fr}}.

Synthetic GW time series data and posterior sample chains are available upon request.

\begin{deluxetable*}{rr|rrrrrrr|rr|rrr|rrr}
\tablecaption{\label{table:2015-sim}Simulated BNS Signals for 2015 Scenario.}
\tablehead{
    \colhead{Event} &
    \colhead{sim} &
    \colhead{\,} &
    \multicolumn{5}{c}{Orientation\tablenotemark{d}} &
    \colhead{$d$} &
    \multicolumn{2}{c}{Masses ($M_\odot$)} &
    \multicolumn{3}{c}{Spin 1} &
    \multicolumn{3}{c}{Spin 2} \\
    \colhead{ID\tablenotemark{a}} &
    \colhead{ID\tablenotemark{b}} &
    \colhead{MJD\tablenotemark{c}} &
    \colhead{$\alpha$} &
    \colhead{$\delta$} &
    \colhead{$\iota$} &
    \colhead{$\psi$} &
    \colhead{$\phi_c$} &
    \colhead{(Mpc)} &
    \colhead{$m_1$} &
    \colhead{$m_2$} &
    \colhead{$S_1^x$} &
    \colhead{$S_1^y$} &
    \colhead{$S_1^z$} &
    \colhead{$S_2^x$} &
    \colhead{$S_2^y$} &
    \colhead{$S_2^z$}
}
\startdata
18951&10807&55442.25864&137.8&-39.9&139&43&42&75&1.40&1.51&-0.01&-0.01&-0.04&-0.05&-0.01&+0.01\\
20342&21002&55454.76654&19.8&-23.7&145&197&145&75&1.34&1.48&-0.03&+0.01&-0.03&-0.01&+0.02&-0.01\\

\dots & \dots & \dots & \dots & \dots & \dots & \dots & \dots & \dots & \dots & \dots & \dots & \dots & \dots & \dots & \dots & \dots
\enddata
\tablenotetext{a}{Identifier for detection candidate. This is the same value as the \texttt{coinc\_event\_id} column in the \textsc{gstlal} output database and the \texttt{OBJECT} cards in sky map FITS headers, with the \texttt{coinc\_event:coinc\_event\_id:} prefix stripped.}
\tablenotetext{b}{Identifier for simulated signal. This is the same value as the \texttt{simulation\_id} column in the \textsc{gstlal} output database, with the \texttt{sim\_inspiral:simulation\_id:} prefix stripped.}
\tablenotetext{c}{Time of arrival at geocenter of GWs from last stable orbit.}
\tablenotetext{d}{$\alpha$: RA, $\delta$: Dec (J2000), $\iota$: binary orbital inclination angle, $\psi$: polarization angle \citep[Appendix B]{ExcessPower}, $\phi_c$: orbital phase at coalescence.}
\tablecomments{(This table is available in its entirety in a machine-readable form in the online journal. A portion is shown here for guidance regarding its form and content.)}
\end{deluxetable*}

\begin{deluxetable*}{rr|r|rrr|rr|rrr|rrr}
\tablecaption{\label{table:2015-coinc}Detections and Sky Localization Areas for 2015 Scenario.}
\tablehead{
    \colhead{Event} &
    \colhead{sim} &
    \colhead{\,} &
    \multicolumn{3}{c}{SNR} &
    \multicolumn{2}{c}{Masses\tablenotemark{b}} &
    \multicolumn{3}{c}{\textsc{bayestar}} &
    \multicolumn{3}{c}{\textsc{lalinference\_nest}} \\
    \colhead{ID} &
    \colhead{ID} &
    \colhead{Network} &
    \colhead{Net\tablenotemark{a}} &
    \colhead{H} &
    \colhead{L} &
    \colhead{$m_1$} &
    \colhead{$m_2$} &
    \colhead{50\%} &
    \colhead{90\%} &
    \colhead{\footnotesize Searched} &
    \colhead{50\%} &
    \colhead{90\%} &
    \colhead{\footnotesize Searched}
}
\startdata
18951&10807&HL&15.0&10.3&10.9&1.67&1.27&159&630&127&158&683&81.2\\
20342&21002&HL&12.7&7.3&10.3&1.59&1.25&126&526&16.9&168&618&12.3\\

\dots & \dots & \dots & \dots & \dots & \dots & \dots & \dots & \dots & \dots & \dots & \dots & \dots & \dots
\enddata
\tablenotetext{a}{Network SNR, or root-sum-squared SNR over all detectors.}
\tablenotetext{b}{Maximum likelihood estimate of masses as reported by \textsc{gstlal}.}
\tablecomments{(This table is available in its entirety in a machine-readable form in the online journal. A portion is shown here for guidance regarding its form and content.)}
\end{deluxetable*}

\begin{deluxetable*}{rr|rrrrrrr|rr|rrr|rrr}
\tablecaption{\label{table:2016-sim}Simulated BNS Signals for 2016 Scenario.}
\tablehead{
    \colhead{Event} &
    \colhead{sim} &
    \colhead{\,} &
    \multicolumn{5}{c}{Orientation} &
    \colhead{$d$} &
    \multicolumn{2}{c}{Masses ($M_\odot$)} &
    \multicolumn{3}{c}{Spin 1} &
    \multicolumn{3}{c}{Spin 2} \\
    \colhead{ID} &
    \colhead{ID} &
    \colhead{MJD} &
    \colhead{$\alpha$} &
    \colhead{$\delta$} &
    \colhead{$\iota$} &
    \colhead{$\psi$} &
    \colhead{$\phi_c$} &
    \colhead{(Mpc)} &
    \colhead{$m_1$} &
    \colhead{$m_2$} &
    \colhead{$S_1^x$} &
    \colhead{$S_1^y$} &
    \colhead{$S_1^z$} &
    \colhead{$S_2^x$} &
    \colhead{$S_2^y$} &
    \colhead{$S_2^z$}
}
\startdata
655803&45345&55484.63177&79.2&+5.0&121&321&69&66&1.60&1.29&+0.00&+0.00&-0.00&+0.00&+0.00&-0.00\\
821759&8914&55439.93634&18.3&-15.1&158&257&230&187&1.60&1.45&-0.00&+0.02&-0.01&+0.04&+0.03&-0.02\\

\dots & \dots & \dots & \dots & \dots & \dots & \dots & \dots & \dots & \dots & \dots & \dots & \dots & \dots & \dots & \dots & \dots
\enddata
\tablecomments{(This table is available in its entirety in a machine-readable form in the online journal. A portion is shown here for guidance regarding its form and content.)}
\end{deluxetable*}

\begin{deluxetable*}{rr|r|rrrr|rr|rrr|rrr}
\tablecaption{\label{table:2016-coinc}Detections and Sky Localization Areas for 2016 Scenario.}
\tablehead{
    \colhead{Event} &
    \colhead{sim} &
    \colhead{\,} &
    \multicolumn{4}{c}{SNR} &
    \multicolumn{2}{c}{Masses} &
    \multicolumn{3}{c}{\textsc{bayestar}} &
    \multicolumn{3}{c}{\textsc{lalinference\_mcmc}} \\
    \colhead{ID} &
    \colhead{ID} &
    \colhead{Network} &
    \colhead{Net} &
    \colhead{H\tablenotemark{a}} &
    \colhead{L\tablenotemark{a}} &
    \colhead{V\tablenotemark{a}} &
    \colhead{$m_1$} &
    \colhead{$m_2$} &
    \colhead{50\%} &
    \colhead{90\%} &
    \colhead{\footnotesize Searched} &
    \colhead{50\%} &
    \colhead{90\%} &
    \colhead{\footnotesize Searched}
}
\startdata
655803&45345&HV&12.2&11.5&&4.2&1.52&1.35&478&4570&65.5&304&3960&20.6\\
821759&8914&HLV&13.4&8.5&10.4&&1.57&1.47&336&1070&473&91.0&515&93.8\\

\dots & \dots & \dots & \dots & \dots & \dots & \dots & \dots & \dots & \dots & \dots & \dots & \dots & \dots & \dots
\enddata
\tablecomments{(This table is available in its entirety in a machine-readable form in the online journal. A portion is shown here for guidance regarding its form and content.)}
\tablenotetext{a}{Blank if $\mathrm{SNR} < 4$ or detector is not online.}
\end{deluxetable*}

\bibliographystyle{apj}
\bibliography{ms}

\end{document}